\documentclass[
  journal=pasa,
  manuscript=research-article,
  year=2024,
  volume=37,
]{cup-journal}

\usepackage{amsmath}
\usepackage{amssymb}
\usepackage[nopatch]{microtype}
\usepackage{booktabs}
\usepackage[acronym]{glossaries-extra}
\usepackage{natbib}
\usepackage{upgreek}
\usepackage{xspace}
\usepackage[normalem]{ulem}
\usepackage{caption}
\usepackage{subcaption}
\usepackage{nicefrac}
\usepackage{xfrac}
\usepackage{rotating}
\usepackage{tabularray}
\usepackage{orcidlink}
\usepackage[textsize=footnotesize]{todonotes}
\setuptodonotes{inline}

\usepackage{tikz}
\usetikzlibrary{shapes.geometric, arrows}
\tikzstyle{startstop} = [rectangle, rounded corners, minimum width=2cm, minimum height=1cm, text centered, draw=black]
\tikzstyle{process} = [rectangle, minimum width=2cm, minimum height=1cm, text centered, draw=black]
\tikzstyle{decision} = [diamond, minimum width=2cm, minimum height=1cm, text centered, draw=black]
\tikzstyle{arrow} = [thick,->,>=stealth]

\usepackage{listings}
\definecolor{codegreen}{rgb}{0,0.6,0}
\definecolor{codegray}{rgb}{0.5,0.5,0.5}
\definecolor{codepurple}{HTML}{C42043}
\definecolor{backcolour}{HTML}{F2F2F2}
\definecolor{bookColor}{cmyk}{0,0,0,0.90}
\lstdefinestyle{sqlstyle}{
        backgroundcolor=\color{backcolour},   
        commentstyle=\color{codegreen},
        keywordstyle=\color{codepurple},
        numberstyle=\footnotesize\color{codegray},
        stringstyle=\color{codepurple},
        basicstyle=\footnotesize\ttfamily,
        breakatwhitespace=false,
        breaklines=true,
        captionpos=b,
        keepspaces=true,
        numbers=left,
        numbersep=5pt,
        showspaces=false,
        showstringspaces=false,
        showtabs=false
}
\lstdefinestyle{pythonstyle}{
    keywordstyle=\color{DarkOrchid3},
    commentstyle={\color{Green4}\itshape\sffamily},
    numberstyle=\tiny\color{Ivory4},
    stringstyle=\color{Purple1},
    basicstyle=\sffamily\footnotesize,
    breakatwhitespace=false,
    breaklines=true,
    captionpos=b,
    keepspaces=true,
    numbers=left,
    numbersep=5pt,
    showspaces=false,
    showstringspaces=false,
    showtabs=false,
    tabsize=2
}

\graphicspath{ {./images/} }

\usepackage{hyperref}
\hypersetup{
    colorlinks=True,
    pdftitle={The MAGPI Survey: Insights into the Lyman-alpha line widths and the size of ionized bubbles at the edge of cosmic reionization - Tamal Mukherjee},
    pdfauthor={Tamal Mukherjee, Macquarie University},
    linkcolor=black,
    urlcolor=blue,
    citecolor=blue
    }

\setabbreviationstyle[acronym]{long-short}
\glsdisablehyper

\newacronym{nir}{near-IR}{near-infrared}
\newacronym{sso}{SSO}{Siding Springs Observatory}
\newacronym{rsaa}{RSAA}{Research School of Astronomy and Astrophysics}
\newacronym{sami}{SAMI}{Sydney-AAO Multi-object Integral field spectrograph}
\newacronym{aat}{AAT}{Anglo-Australian Telescope}
\newacronym{ifs}{IFS}{integral-field spectroscopy}
\newacronym{uv}{UV}{ultraviolet}
\newacronym{sed}{SED}{spectral energy distribution}
\newacronym{ir}{IR}{infrared}
\newacronym{sfr}{SFR}{star formation rate}
\newacronym{mw}{MW}{Milky Way}
\newacronym{ifu}{IFU}{integral field unit}
\newacronym{gama}{GAMA}{Galaxy and Mass Assembly}
\newacronym{dr3}{DR3}{third data release}
\newacronym{snr}{SNR}{signal-to-noise ratio}
\newacronym{agn}{AGN}{active galactic nuclei}
\newacronym{dig}{DIG}{diffuse ionised gas}
\newacronym{magpi}{MAGPI}{Middle Ages Galaxy Properties with Integral field spectroscopy}
\newacronym{manga}{MaNGA}{Mapping Nearby Galaxies at Apache Point Observatory}
\newacronym{3dhst}{3D-HST}{Three-Dimensional Hubble Space Telescope}
\newacronym{califa}{CALIFA}{Calar Alto Legacy Integral Field Area}
\newacronym{muse}{MUSE}{Multi Unit Spectroscopic Explorer}
\newacronym{hudf}{HUDF}{Hubble Ultra-Deep Field}
\newacronym{glao}{GLAO}{ground-layer adaptive optics}
\newacronym{lgs}{LGS}{laser guide star}
\newacronym{pmas}{PMAS}{Potsdam Multi Aperture Spectrograph}
\newacronym{jades}{JADES}{JWST Advanced Deep Extragalactic Survey}
\newacronym{ism}{ISM}{interstellar medium}
\newacronym{cirs}{CIRS}{Cluster Infall Regions in the SDSS}
\newacronym{2dfgrs}{2dFGRS}{2dF Galaxy Redshift Survey}
\newacronym{ssfr}{sSFR}{specific star formation rate}
\newacronym{dar}{DAR}{differential atmospheric refraction}
\newacronym{sps}{SPS}{stellar population synthesis}
\newacronym{liner}{LINER}{low-ionisation nuclear emission-line region}
\newacronym{lae}{LAE}{Lyman-$\upalpha$ emitter}
\newacronym{jwst}{JWST}{James Webb Space Telescope}
\newacronym{gist}{GIST}{Galaxy IFU Spectroscopy Tool}

\def\ha{H$\alpha$}

\def\lya{Ly$\alpha$}

\def\oiii{[O~{\sc iii}]}

\def\hi{H~{\sc i}}
\def\hii{H~{\sc ii}}

\def\ciii{C~{\sc iii}]}
\def\civ{C~{\sc iv}}

\def\mgii{Mg~{\sc ii}}

\title{The MAGPI Survey: Insights into the \lya\ line widths and the size of ionized bubbles at the edge of cosmic reionization}

\author{T. Mukherjee
\textsuperscript{\orcidlink{0009-0004-7639-869X}}}
\affiliation{School of Mathematical and Physical Sciences, Macquarie University, NSW 2109, Australia}
\alsoaffiliation{ARC Centre of Excellence for All Sky Astrophysics in 3 Dimensions (ASTRO 3D)}
\alsoaffiliation{Astrophysics and Space Technologies Research Centre, Macquarie University, Sydney, NSW 2109, Australia}

\email[T. Mukherjee]{tamal.mukherjee@hdr.mq.edu.au}

\author{T. Zafar
\textsuperscript{\orcidlink{0000-0003-3935-7018}}}
\affiliation{School of Mathematical and Physical Sciences, Macquarie University, NSW 2109, Australia}
\alsoaffiliation{ARC Centre of Excellence for All Sky Astrophysics in 3 Dimensions (ASTRO 3D)}
\alsoaffiliation{Astrophysics and Space Technologies Research Centre, Macquarie University, Sydney, NSW 2109, Australia}

\author{T. Nanayakkara
\textsuperscript{\orcidlink{0000-0003-2804-0648}}}
\affiliation{Centre for Astrophysics and Supercomputing, Swinburne University of Technology, PO Box 218, Hawthorn 3122, VIC, Australia}
\alsoaffiliation{ARC Centre of Excellence for All Sky Astrophysics in 3 Dimensions (ASTRO 3D)}

\author{A. Gupta
\textsuperscript{\orcidlink{0000-0002-8984-3666}}}
\affiliation{International Centre for Radio Astronomy Research (ICRAR), Curtin University, Bentley, WA, Australia
}
\alsoaffiliation{ARC Centre of Excellence for All Sky Astrophysics in 3 Dimensions (ASTRO 3D)}

\author{S. Gurung-Lopez
\textsuperscript{\orcidlink{0000-0001-9333-8470}}}
\affiliation{Observatori Astron\`omic de la Universitat de Val\`encia, Ed. Instituts d’Investigaci\'o, Parc Cient\'ific. C/ Catedr\'atico Jos\'e Beltr\'an, n2, 46980 Paterna, Valencia, Spain
}
\alsoaffiliation{Departament d’Astronomia i Astrof\'isica, Universitat de Val\`encia, 46100 Burjassot, Spain}

\author{A. Battisti
\textsuperscript{\orcidlink{0000-0003-4569-2285}}}
\affiliation{Research School of Astronomy and Astrophysics, Australian National University, Canberra, ACT 2611, Australia
}
\alsoaffiliation{ARC Centre of Excellence for All Sky Astrophysics in 3 Dimensions (ASTRO 3D)}

\author{E. Wisnioski
\textsuperscript{\orcidlink{0000-0003-1657-7878}}}
\affiliation{Research School of Astronomy and Astrophysics, Australian National University, Canberra, ACT 2611, Australia
}
\alsoaffiliation{ARC Centre of Excellence for All Sky Astrophysics in 3 Dimensions (ASTRO 3D)}

\author{C. Foster
\textsuperscript{\orcidlink{0000-0003-0247-1204}}}
\affiliation{School of Physics, University of New South Wales, Sydney, NSW 2052, Australia}
\alsoaffiliation{ARC Centre of Excellence for All Sky Astrophysics in 3 Dimensions (ASTRO 3D)}

\author{J. T. Mendel
\textsuperscript{\orcidlink{0000-0002-6327-9147}}}
\affiliation{Research School of Astronomy and Astrophysics, Australian National University, Canberra, ACT 2611, Australia
}
\alsoaffiliation{ARC Centre of Excellence for All Sky Astrophysics in 3 Dimensions (ASTRO 3D)}

\author{K. E. Harborne
\textsuperscript{\orcidlink{0000-0002-2043-7985}}}
\affiliation{International Centre for Radio Astronomy Research (ICRAR), The University of Western Australia, Crawley, WA 6009, Australia
}
\alsoaffiliation{ARC Centre of Excellence for All Sky Astrophysics in 3 Dimensions (ASTRO 3D)}

\author{C. D. P. Lagos
\textsuperscript{\orcidlink{0000-0003-3021-8564}}}
\affiliation{International Centre for Radio Astronomy Research (ICRAR), The University of Western Australia, Crawley, WA 6009, Australia
}
\alsoaffiliation{ARC Centre of Excellence for All Sky Astrophysics in 3 Dimensions (ASTRO 3D)}

\author{T. Kodama
\textsuperscript{\orcidlink{0000-0002-2993-1576}}}
\affiliation{Astronomical Institute, Tohoku University, 6-3 Aramaki, Aoba-ku, Sendai 980-8578, Japan}

\author{S. M. Croom
\textsuperscript{\orcidlink{0000-0003-2880-9197}}}
\affiliation{Sydney Institute for Astronomy, School of Physics, University of Sydney, NSW 2006, Australia
}
\alsoaffiliation{ARC Centre of Excellence for All Sky Astrophysics in 3 Dimensions (ASTRO 3D)}

\author{S. Thater
\textsuperscript{\orcidlink{0000-0003-1820-2041}}}
\affiliation{Department of Astrophysics, University of
Vienna, Türkenschanzstraße 17, 1180 Vienna}

\author{J. Webb
\textsuperscript{\orcidlink{0009-0006-7339-1112}}}
\affiliation{School of Mathematical and Physical Sciences, Macquarie University, NSW 2109, Australia}
\alsoaffiliation{Astrophysics and Space Technologies Research Centre, Macquarie University, Sydney, NSW 2109, Australia}

\author{S. Barsanti
\textsuperscript{\orcidlink{0000-0002-9332-5386}}}
\affiliation{Research School of Astronomy and Astrophysics, Australian National University, Canberra, ACT 2611, Australia}
\alsoaffiliation{ARC Centre of Excellence for All Sky Astrophysics in 3 Dimensions (ASTRO 3D)}
\alsoaffiliation{Sydney Institute for Astronomy, School of Physics, University of Sydney, NSW 2006, Australia}

\author{S. M. Sweet
\textsuperscript{\orcidlink{0000-0002-1576-2505}}}
\affiliation{School of Mathematics and Physics, University of Queensland, Brisbane, QLD 4072, Australia}
\alsoaffiliation{ARC Centre of Excellence for All Sky Astrophysics in 3 Dimensions (ASTRO 3D)}

\author{J. Prathap
\textsuperscript{\orcidlink{0009-0004-0251-2672}}}
\affiliation{School of Mathematical and Physical Sciences, Macquarie University, NSW 2109, Australia}
\alsoaffiliation{ARC Centre of Excellence for All Sky Astrophysics in 3 Dimensions (ASTRO 3D)}
\alsoaffiliation{Astrophysics and Space Technologies Research Centre, Macquarie University, Sydney, NSW 2109, Australia}

\author{L. M. Valenzuela
\textsuperscript{\orcidlink{0000-0002-7972-9675}}}
\affiliation{Universitäts-Sternwarte, Fakultät für Physik, 
Ludwig-Maximilians-Universität München, Scheinerstr. 1, 81679 München, 
Germany}

\author{A. Mailvaganam
\textsuperscript{\orcidlink{0009-0003-1221-1630}}}
\affiliation{School of Mathematical and Physical Sciences, Macquarie University, NSW 2109, Australia}
\alsoaffiliation{ARC Centre of Excellence for All Sky Astrophysics in 3 Dimensions (ASTRO 3D)}
\alsoaffiliation{Astrophysics and Space Technologies Research Centre, Macquarie University, Sydney, NSW 2109, Australia}

\author{J. L. Carrillo Martinez}
\affiliation{School of Mathematical and Physical Sciences, Macquarie University, NSW 2109, Australia}
\alsoaffiliation{ARC Centre of Excellence for All Sky Astrophysics in 3 Dimensions (ASTRO 3D)}
\alsoaffiliation{Astrophysics and Space Technologies Research Centre, Macquarie University, Sydney, NSW 2109, Australia}

\keywords{cosmology: dark ages, reionization, first stars – galaxies: evolution, high redshift, intergalactic medium} %% First letter not capped

\begin{document}

\begin{abstract}

We present spectroscopic properties of 22 \lya\ emitters (LAEs) at $z = 5.5 - 6.6$ with \lya\ luminosity $\mathrm{log}( L_{\mathrm{Ly}\alpha} \, [\mathrm{erg} \, \mathrm{s}^{-1}]) = 42.4 - 43.5 $, obtained using VLT/MUSE as part of the Middle Ages Galaxy Properties with Integral Field Spectroscopy (MAGPI) survey. Additionally, we incorporate broad-band photometric data from the Subaru Hyper Suprime-Cam (HSC) Wide layer for $17$ LAEs in our sample. The HSC-$y$ band magnitudes show that our LAEs are UV-bright, with rest-frame absolute UV magnitudes $ -19.74 \leq \mathrm{M}_{\mathrm{UV}} \leq -23.27$. We find that the \lya\ line width increases with \lya\ luminosity, and this trend becomes more prominent at $z > 6$ where \lya\ lines become significantly broadened ($\gtrsim 260 \, \mathrm{km}\, \mathrm{s}^{-1}$) at luminosities $\mathrm{log}( L_{\mathrm{Ly}\alpha} \, [\mathrm{erg} \, \mathrm{s}^{-1}]) > 43 $. This broadening is consistent with previous studies, suggesting that these sources are located inside larger ionized bubbles. We observe a slightly elevated ionizing photon production efficiency estimated for LAEs at $z > 6$, which indicates that younger galaxies could be producing more ionizing photons per UV luminosity. A tentative anti-correlation between ionizing photon production efficiency and \lya\ rest-frame equivalent width is noticed, which could indicate a time delay between production and escape of ionizing photon primarily due to supernovae activity. Furthermore, we find a positive correlation between radius of ionized regions and \lya\ line width, which again suggests that large ionized bubbles are created around these LAEs, which are allowing them to self-shield from the scattering effects of the intergalactic medium. We also detect two very closely separated LAEs at $z = 6.046$ (projected spatial distance between the cores is 15.92 kpc). This is the LAE pair with the smallest separation ever discovered in the reionization epoch. The size of their respective bubbles suggests that they likely sit inside a common large ionized region. Such a closely-separated LAE pair increases the size of ionized bubble, potentially allowing a boosted transmission of \lya\ through neutral intergalactic medium and also supports an accelerated reionization scenario.  

%A recent study suggests that \lya\ emitters (LAEs) during the reionization era with observed \lya\ luminosities $\mathrm{log} \, L (\mathrm{Ly}\alpha) \gtrsim 43.25$ $\mathrm{ergs} \, \mathrm{s}^{-1}$ mark the ionized regions. We report the detection of three very luminous LAEs - MAGPI Redshift 5 or MR5 (at $z=5.5$), MAGPI Redshift 6 or MR6 (at $z=6.1$) and MAGPI Redshift 7 or MR7 (at $z=6.6$)  with  \lya\ luminosities $\mathrm{log} \, L (\mathrm{Ly}\alpha)=43.32 - 43.5$ $\mathrm{ergs} \, \mathrm{s}^{-1}$, in the VLT/MUSE data obtained as part of the Middle Ages Galaxy Properties with Integral Field Spectroscopy (MAGPI) survey. MR5 and MR7 are also detected in the broadband filters g, r, i, z and y on Hyper Suprime-Cam (HSC) on the Subaru telescope. 
 
\end{abstract}

%\noindent Lorem ipsum dolor sit amet, consectetur adipiscing elit, sed do eiusmod tempor incididunt ut labore et dolore magna aliqua. 

\section{Introduction}\label{sec:intro}

Cosmic reionization, a pivotal epoch in the history of the Universe, marks the last phase transition of the Universe when neutral hydrogen (\hi) in the intergalactic medium (IGM) became fully ionized, ending the cosmic ``Dark Ages." However, the precise timing of reionization and sources capable of emitting sufficient ionizing photons remain subjects of active debate till date. Previously, it was believed that reionization was largely complete by $z\sim 6$ \citep{Fan06}. However, recent studies suggest a relatively late end of reionization at $z\sim 5.3 - 5.5$ \citep{Becker15, Kulkarni19, Cain21, Bosman22}. A common belief is that galaxies which are faint in intrinsic ultraviolet (UV) radiation are the primary contributors of reionization, typically releasing about 10 \% of their Lyman continuum (LyC) photons \citep[see][]{Finkelstein19, Dayal20}. However, to explain the relatively rapid decrease in the neutral IGM fraction in later epochs, it is possible that more infrequent luminous sources might also have played a significant role \citep[see][]{Naidu20}. 

Due to the attenuation of UV photons below the Lyman break by the increasingly neutral IGM at $z > 4$ \citep[e.g.][]{Inoue14, Steidel18}, direct observation of LyC photons is almost impossible. Reionization models suggest that a minimum escape fraction of LyC photons, $f^{\mathrm{LyC}}_{\mathrm{esc}} \gtrsim 10$ \% is required to complete reionization \citep{Robertson15, Finkelstein19}. Therefore, it is crucial to comprehend how LyC photons escape into the IGM and subsequently ionize it during the epoch of reionization (EoR). Several studies have attempted to make connection between $f^{\mathrm{LyC}}_{\mathrm{esc}}$ and nebular emission line features such as \oiii, \civ, \ciii\, \mgii\, etc \citep[see,][]{Izotov20, Nakajima20, Schaerer22, Katz22, Xu22, Mascia23, Choustikov24}. 
%However, these studies are limited to $z \lesssim 5$.    

The \lya\ emission line of atomic hydrogen has been identified as the most reliable indirect tracer of LyC leakage and is used as one of the promising probes of the EoR \citep{Kakiichi16, Laursen19, Tang23}. A rapid decline in the fraction of \lya\ emitting galaxies (LAEs) towards higher redshifts ($z > 5$) has been interpreted as a rapid escalation in the \hi\ fraction with increasing redshift \citep{Pentericci11, Tilvi14, Stark17, Hoag19, Whitler20, Jones24, Nakane24, Napolitano24, Tang24}. Both the observed intensity and shape of the \lya\ line offer delicate insights into the proportion of \hi\ within the IGM \citep{Robert10}. The \lya\ rest-frame equivalent width ($\mathrm{EW}_0$) has been identified as an excellent indicator of \lya\ escape fraction ($f^{\mathrm{Ly}\alpha}_{\mathrm{esc}}$; see \citealt{Matthee17, Begley24, Tang24b}). The separation between the blue and red peaks in the double-peak emission can be used to infer \hi\ column densities ($N_{\mathrm{HI}}$), consequently, the escape of LyC photons \citep[see,][]{Verhamme15, Verhamme17, Izotov18, Naidu22}. Furthermore, the detection of a stronger blue-peak profile also indicates a very low column density channel of \hi\ that can leak LyC photons \citep[see,][]{Furtak22, Mukherjee23}. However, the blue-peak is expected to be scattered away by the neutral IGM at $z > 5$ \citep{Hu10, Hayes21}, leaving only a single-peak red-skewed profile.  

Narrow-band (NB) surveys have discovered a substantial sample of LAE at $z = 5.7$ and $z = 6.6$ \citep[see][]{Hu10, Matthee15, Santos16, Bagley17, Konno18, Taylor20, Taylor21}, and several LAE have also been detected at $z = 6.9$ \citep{Hu17} and $z = 7.3$ \citep{Konno14}. Recent data from the James Webb Space Telescope (JWST) has also contributed to the unprecedented characterization of LAEs throughout the EoR \citep{Tang23, Jung24, Witten24}. The advent of giant imagers, such as Subaru/ Hyper Suprime-Cam \citep[HSC;][]{Miyazaki18}, has allowed detections of several rare ultra-luminous LAEs (ULLAEs) with $\mathrm{log}( L_{\mathrm{Ly}\alpha} \, [\mathrm{erg} \, \mathrm{s}^{-1}]) > 43.5 $ \citep[see,][]{Songaila22}, including the detections of extremely rare double-peaked LAEs with a blue-wing \citep[see][]{Hu16, Songaila18, Meyer21}. 

The Multi Unit Spectroscopic Explorer \citep[see,][]{Bacon10} on the Very Large Telescope (VLT) has been instrumental in identifying faint LAEs during the EoR (up to $z \sim 6.6$) by providing deep, high-resolution spectral data across a wide field of view, without any redshift restrictions of NB imaging \citep[see,][]{Hashimoto17, Urrutia19, Kerutt22, Bacon23}. 

Reionization is known to be an inhomogeneous process \citep{Pentericci14, Becker15, Bosman22}, indicating that galaxies in denser regions are likely to create the first 'ionized bubbles' in the Universe \citep{Mason18a, Endsley1, Jung22a, Endsley22, Whitler24}, which then preferentially emit \lya\ radiation once they reach a significant size. As a result, the most distant LAEs are vital for observing and mapping the reionization process. Strong \lya\ emission at $z \gtrsim 6$ often indicates the presence of large ionized bubbles in an otherwise neutral IGM, providing direct observational insights into the reionized regions of the early Universe. When \lya\ photons are emitted from galaxies located within large ionized bubbles, they undergo cosmological redshifting far into the damping wing before encountering neutral hydrogen. As a result, they transmit more effectively through IGM \citep{Weinberger18, MG20, Smith22, Qin22}. Recent studies have found that \lya\ line width increases with luminosity and this trend becomes more prominent at $z = 6.6$ \citep[see,][]{Matthee17, Songaila24}, where higher-luminosity LAEs with $\mathrm{log}( L_{\mathrm{Ly}\alpha} \, [\mathrm{erg} \, \mathrm{s}^{-1}]) \gtrsim 43.25 $ show significantly broad \lya\ lines \citep{Songaila24} . At higher redshifts, the IGM becomes more neutral, increasing \lya\ line scattering and hence narrowing of the lines is expected. The lack of this effect in brighter LAEs suggests that they reside in more ionized regions, shielding themselves from scattering. 

One of the key components in determining the ionizing photon budget is the ionizing photon production efficiency ($\xi_{\mathrm{ion}}$), which is defined as the ratio between the production rate of ionizing photons over the non-ionizing UV luminosity density. It has been found that as we delve deeper into the universe's history, young galaxies seem to appear more efficient in producing ionizing photons \citep[see,][]{Bouwens16, Endsley2, Lyon23, Simmonds23, Tang23}. Moreover, UV-faint galaxies with \lya\ emission are found to have enhanced $\xi_{\mathrm{ion}}$ \citep{Maseda20, Ning23, Saxena24, Lin24} at $z \sim 6$, making LAEs in the reionization era the most exciting sources for studying and constraining reionization.   

In this paper, we present spectroscopic properties of $22$ new LAEs at the end of reionization ($z = 5.5 - 6.6$) found in the MUSE data obtained as a part of the Middle Ages Galaxy Properties with Integral Field Spectroscopy (MAGPI) survey \citep{Foster21}. We restrict this current study to redshifts of $z \gtrsim 5.5$, considering the global neutral hydrogen fraction approaches $x_{HI} \sim 0$ at around $z \sim 5.5$ in `late’ reionization scenarios \citep{Kulkarni19, Bosman22}. We constrain the evolution of \lya\ line width as a function of \lya\ luminosity up to $z \sim 7$. Using spectroscopic information of these LAEs, along with the Subaru Hyper-Suprime Cam (HSC) optical photometric information, we estimate their ionising photon contribution toward the global reionization budget. We also estimate the size of ionized bubbles around these LAEs to investigate the mechanism leading to the visibility of strong \lya\ emission even beyond $z > 6$. 

The layout of this paper is as follows: In \S \ref{sec:obs}, we describe the reduction and selection of MUSE data of potential LAE candidates along with the extraction of HSC photometric magnitudes. \S \ref{sec:analyses} explores the data analyses: \lya\ line fitting and spectroscopic and photometric measurements. \S \ref{sec:results} presents results on the evolution of the \lya\ line widths and give an insight into the potential ionized bubbles surrounding these LAEs and also discusses these findings in context of previous works. The main conclusions and summary of this study are presented in \S \ref{sec:summary}. Throughout this paper, we assume a standard flat $\Lambda$CDM cosmology with parameters $H_0$= 70 $\mathrm{km} \, \mathrm{s}^{-1} \mathrm{Mpc}^{-1}$, $\Omega_{\mathrm{m}}$ = $0.3$ and  $\Omega_{\Lambda}$ = $0.7$.

\section{Observations and data}\label{sec:obs}
\subsection{MUSE spectroscopic data}
The MAGPI survey \footnote{Based on observations obtained using MUSE instrument at VLT of the European Southern Observatory (ESO), Paranal, Chile (ESO program ID $1104.\mathrm{B}-0536$)} is an ongoing Large Program on the VLT/MUSE, targeting $56$ fields from the Galaxy and Mass Assembly \citep[GAMA;][]{Driver11} G12, G15 and G23 fields. MAGPI also includes archival observations of legacy fields Abell $370$ and Abell $2744$. The survey targets a total of $60$ primary galaxies with stellar masses $M_{*} > 7 \times10^{10}$\,$M_\odot$ and $\sim 100$ satellite galaxies with  $M_{*} > 10^{9}$\,$M_\odot$. The primary objective of MAGPI is to conduct a detailed spatially resolved spectroscopic analysis of stars and ionized gas within $0.25< z <0.35$ galaxies \citep[see][]{Foster21}. Data are taken using the MUSE Wide Field Mode ($1'\times1'$) with a spatial sampling rate of $0.2''$/pixel and the median Full Width at Half Maximum (FWHM) is $0.64''$ in $g$ band, $0.6''$ in $r$ band and $0.55''$ in $i$ band. Each field is observed in six observing blocks, each comprising 2$\times$1320\,s exposures, resulting in a total integration time of $4.4$\,h. The survey primarily employs the nominal mode, providing a wavelength coverage ranging from $4700$\,\AA\ to $9350$\,\AA, with a dispersion of $1.25$\,\AA. Ground-layer adaptive optics (GLAO) is used to correct atmospheric seeing effects, resulting in a gap between $5805$\,\AA\ and $5965$\,\AA\ due to the GALACSI laser notch filter. The depth of MAGPI data allows for the detection of both foreground sources within the Local Universe and distant background sources, including LAEs at $ 2.9 \lesssim z \lesssim 6.6$. 

The raw MUSE data cubes are reduced using \texttt{ Pymusepipe2}\footnote{\href{https://github.com/emsellem/pymusepipe}{https://github.com/emsellem/pymusepipe}}, a Python wrapper for the ESO MUSE reduction pipeline \citep{Weilbacher20}. This pipeline is used to perform the standard bias and overscan subtraction, flat-fielding, wavelength calibration and telluric correction. Additional information about the data reduction process is provided in \cite{Foster21} and it will be presented in a greater detail (Mendel et al. in prep.). \texttt{LSDCat} \footnote{\href{https://bitbucket.org/Knusper2000/lsdcat}{https://bitbucket.org/Knusper2000/lsdcat}} \citep{HW17} was used for the identification of faint sources --- particularly LAEs --- accompanied by both automated and visual inspection. LAEs are confirmed using visual inspections of line-profiles and using \texttt{MARZ}\footnote{\href{https://github.com/Samreay/Marz}{https://github.com/Samreay/Marz}} redshifting software \citep{Hinton16} to rule out interlopers. This search led to the detection of $380$ new LAEs distributed across $35$ MAGPI fields (Mukherjee et al. in prep). Among these, we found $22$ LAEs with $\mathrm{log} \,( L_{\mathrm{Ly}\alpha} \, [\mathrm{erg} \, \mathrm{s}^{-1}]) > 42.4 $ at $5.5 \lesssim z \lesssim 6.6$. %Three of them are very luminous with observed \lya\ luminosities $\mathrm{log} \, L (\mathrm{Ly}\alpha) \gtrsim 43.3$ $\mathrm{ergs} \, \mathrm{s}^{-1}$. 

\begin{table}[hbt!]
\begin{threeparttable}
\caption{HSC photometry of MAGPI LAEs at $z = 5.5 - 6.6$, in order of increasing redshift. HSC-$i$, $z$ and $y$ band AB magnitudes and $2\sigma$ limits (for non-detections) are presented. $\mathrm{M}_{\mathrm{UV}}$ is the rest-frame absolute UV magnitude estimated from the $y$-band magnitudes.}
\label{t1}
\begin{tabular}{lcccc}
\toprule
\headrow MAGPI ID & HSC-$i$ &  HSC-$z$  & HSC-$y$ & $\mathrm{M}_{\mathrm{UV}}$ \\
\midrule
$1507091138$ & $25.52 \pm 0.12$ & $25.10 \, \pm 0.18$ & $24.43 \pm 0.19$ & $-22.20$\\ 
%\midrule
$1527275156$ & $25.29 \pm 0.13$ & $25.34 \, \pm 0.29 $ & $24.46 \pm 0.21$ & $-22.17$\\ 
%\midrule
2310233132 & $-$ & $-$ & $-$ & $-$    \\ 
%\midrule
1527283124 & $26.21 \pm 0.33$ & $26.10 \, \pm 0.45$ & $26.66 \pm 1.08$ & $-19.96$ \\ 
%\midrule
1503111271 & $26.04 \pm 0.24$ & $26.86 \, \pm 0.91$ & $25.82 \pm 0.86$ & $-20.80$  \\ 
%\midrule
1507313178 & $> 27.18$ & $> 26.86$ & $> 25.75$ & $> -20.87$ \\ 
%\midrule
1511268137 & $> 27.55$ & $> 27.49$ & $> 26.21$ & $> -20.41$ \\ 
%\midrule
1207184066 & $-$ & $-$ & $-$ & $-$  \\
%\midrule
2306257117 & $-$ & $-$ & $-$ & $-$  \\ 
%\midrule
1205187075 & $> 28.25$ & $> 29$ & $> 26.40 $ & $> -20.22$
\\
%\midrule
1523134187 & $25.59 \pm 0.19$ & $24.39 \pm 0.32$ & $24.31 \pm 0.33$ & $-22.30$
\\
%\midrule
1507308274 & $> 29$ & $> 27.37$ & $> 25.19$ & $> -21.42$
\\
%\midrule
2310245276 & $-$ & $-$ & $-$ & $-$   \\ 
%\midrule
1204117107 & $26.82 \pm 0.49$ & $25.26 \, \pm 0.25$ & $24.33 \pm 0.19$ & $-22.28$ \\ 
%\midrule
1529110045 & $25.74 \pm 0.13$ & $25.26 \, \pm 0.16$ & $25.60 \pm 0.40$ & $-21.01$ \\ 
1529106057 & $25.93 \pm 0.23$ & $24.49 \, \pm 0.16$ & $ 25.06 \pm 0.35$ & $- 21.55$ \\ 
%\midrule
%\midrule
%1206233254 & xx & xx & xx \\ 
%\midrule
2310222098 & $-$ & $-$ & $-$ & $-$   \\ 
%\midrule
1528094186 & $26.72 \pm 0.42$ & $< 27.61$ & $25.38 \pm 0.5$ & $-21.22$\\ 
%\midrule
1505103163 & $25.85 \pm 0.25$ & $25.84 \, \pm 0.56$ & $23.32 \pm 0.30$ & $-23.27$\\ 
%\midrule
%1523213176 & $0.007 \, \pm 0.003$ & $25.77 \, \pm 0.08$ & 0.24 \\ 
%\midrule
1530068179 & $> 26.17$ & $> 27$ & $> 26.85$ & $> -19.74$\\ 
%\midrule
1528263095 & $> 26.13$ & $> 27.44$ & $> 26.46$ & $> -20.12$ \\
%\midrule
1522272275  & $25.87 \pm 0.22$ & $27.05 \, \pm 1.01$ & $25.25 \pm 0.41$ & $-21.33$\\ 
\bottomrule
\end{tabular}
\end{threeparttable}
\end{table}

\begin{table*}[hbt!]
\begin{threeparttable}
\caption{Properties of $22$ MAGPI LAEs at $5.5 \lesssim z \lesssim 6.6$ in the sample, in order of increasing redshift. Columns are as follows: MAGPI ID; RA: Right Ascension in degrees; DEC: Declination in degrees; $z$: Redshift based on the peak of the \lya\ line; $\mathrm{log}_{10}(L_{\mathrm{Ly}\alpha})$: observed \lya\ luminosity in $\mathrm{erg}\, \mathrm{s}^{-1}$; $\mathrm{FWHM}$: \lya\ line width measured using asymmetric Gaussian fit; $f^{\mathrm{cont}}_{\mathrm{Ly}\alpha}$ $^{*}$: Observed UV-continuum flux-density at the location of \lya\ wavelength, in $10^{-20} \, \mathrm{erg}\, \mathrm{s}^{-1} \mathrm{cm}^{-2}$\AA$^{-1}$; $\mathrm{EW}_{0}$: Spectroscopically measured \lya\ rest-frame equivalent width; $f^{\mathrm{Ly}\alpha}_{\mathrm{esc}}$: \lya\ escape fraction; $\mathrm{log}(\xi_{\mathrm{ion}})$: Ionizing photon production efficiency in $\mathrm{Hz}\, \mathrm{erg}^{-1}$; $R_{\mathrm{B}}$: radius of bubble ionized by LAE itself, in pMpc. }
\label{t2}
\begin{tabular}{lccccccccccc}
\toprule
\headrow MAGPI ID & RA & DEC  & $z$ & $\mathrm{log}_{10}(L_{\mathrm{Ly}\alpha})$ & $\mathrm{FWHM}$ & $f^{\mathrm{cont}}_{\mathrm{Ly}\alpha}$ & $\mathrm{EW}_{0}$  & $f^{\mathrm{Ly}\alpha}_{\mathrm{esc}}$ &  $\mathrm{log}(\xi_{\mathrm{ion}})$ & $R_{\mathrm{B}}$\\
 & [deg] & [deg] &  &  $[\mathrm{erg}\, \mathrm{s}^{-1}]$ & $[\mathrm{km}\,\mathrm{s}^{-1}]$ & * & [\AA] &  &  $[\mathrm{Hz}\, \mathrm{erg}^{-1}]$ & [pMpc]\\
\midrule
$1507091138$ & $215.6265$ & $0.4041$ & $5.4955$ & $42.77 \pm 0.04$  &  $235.45 \,\pm 24.65$ & $12.09$ & $23.18$ $\pm 3.28$ & $0.087^{+ 0.019}_{-0.017}$ & $25.26 \pm 0.03 $& $0.55 \pm 0.02$ \\ 
%\midrule
$1527275156$ & $220.0398$ & $-0.6588$ & $5.5010$ & $43.50\pm 0.01$  &  $274.29 \,\pm 8.56$ & $19.89$ & $72.24$ $\pm 9.65$ & $0.273^{+ 0.060}_{-0.054}$ & $25.51 \pm 0.01$ & $0.66 \pm 0.04$\\ 
%\midrule
$2310233132$ & $348.3007$ & $-34.0181$ & $5.5267$ & $42.76\pm 0.03$  &  $240.43 \,\pm 15.01$ & $4.05$ & $66.27$ $\pm 12.43$ & $0.251^{+ 0.070}_{-0.062}$ &$-$ & $0.38 \pm 0.03$\\ 
%\midrule
$1527283124$ & $220.0399$ & $-0.6613$ & $5.5393$ & $42.98\pm 0.04$  &  $211.52 \,\pm 17.15$ & $6.28$ & $70.2$ $\pm 17.91$ & $0.266^{+ 0.094}_{-0.083}$ & $25.88 \pm 0.34$ & $0.44 \pm 0.05$\\ 
%\midrule
$1503111271$ & $213.6237$ & $-0.4111$ & $5.6483$ & $42.97\pm 0.03$  &  $252.83 \,\pm 27.29$ & $20.88$ & $19.74$ $\pm 4.71$ & $0.074^{+ 0.024}_{-0.021}$ & $26.08 \pm 0.26$ & $0.65 \pm 0.07$\\ 
%\midrule
$1507313178$ & $215.6141$ & $0.4064$ & $5.6570$ & $42.87 \pm 0.03$  &  $272.68 \,\pm 15.15$ & $9.45$ & $34.50$ $\pm 7.16$ & $0.130^{+ 0.039}_{-0.034}$ & $< 25.72$ & $0.51 \pm 0.04$\\ 
%\midrule
$1511268137$ & $216.5675$ & $1.7274$ & $5.7198$ & $42.55\pm 0.05$ & $175.55 \,\pm 17.04$ & $2.17$ & $69.01$ $\pm 37.78$ & $0.261^{+ 0.175}_{-0.152}$ & $< 25.28$ & $0.31 \pm 0.09$\\ 
%\midrule
$1207184066$ & $182.0006$ & $-2.4909$ & $5.7643$ & $42.40 \pm 0.04$ &  $295.35 \, \pm 32.29$ & $3.27$ & $29.36$ $\pm 8.73$ & $0.110^{+ 0.043}_{-0.038}$ & $-$ & $0.37 \pm 0.04$\\
%\midrule
$2306257117$ & $345.0617$ & $-34.4722$ & $5.7750$ & $42.40\pm 0.06$ &  $213.43 \, \pm 17.21$ & $4.09$ & $25.40$ $\pm 8.09$ & $0.095^{+ 0.040}_{-0.035}$ & $-$ & $0.38 \pm 0.05$
\\ 
%\midrule
$1205187075$ & $178.0804$ & $-0.8336$ & $5.8010$ & $42.47\pm 0.05$ &  $266.47 \, \pm 46.14$ & $3.22$ & $37.10$ $\pm 15.06$ & $0.140^{+ 0.074}_{-0.065}$ & $< 25.54$ & $0.36 \pm 0.06$
\\
%\midrule
$1523134187$ & $219.5444$ & $-1.0999$ & $5.9285$ & $42.65\pm 0.03$ &  $237.48 \, \pm 34.83$ & $3.77$ & $42.10$ $\pm 14.68$ & $0.158^{+ 0.072}_{-0.063}$ & $24.84 \pm 0.01$ & $0.39 \pm 0.06$
\\
%\midrule
$1507308274$ & $215.6144$ & $0.4117$ & $5.9815$ & $42.73\pm 0.03$ &  $256.93 \, \pm 33.13$ & $19.36$ & $10.06$ $\pm 1.59$ & $0.037^{+ 0.008}_{-0.007}$ & $<25.90$ & $0.66 \pm 0.03$
\\
$2310245276$ & $348.2999$ & $-34.0101$ & $6.0390$ & $42.52\pm 0.04$  &  $199.64 \,\pm 20.64$ & $4.96$ & $23.54$ $\pm 6.15$ & $0.088^{+ 0.031}_{-0.027}$ & $-$ & $0.39 \pm 0.07$ \\ 
%\midrule
$1204117107$  & $175.6656$ & $-0.79936$ & $6.0460$ & $43.18\pm 0.02$  &  $367.48 \,\pm 25.83$ & $35.76$ & $12.50$ $\pm 3.15$ & $0.046^{+ 0.016}_{-0.014}$  & $25.91 \pm 0.03$ & $ 0.86 \pm 0.09$\\ 
%\midrule
$1529110045$ & $220.3717$ & $-0.1079$ & $6.0462$ & $42.93\pm 0.03$  &  $197.49 \,\pm 18.16$ & $8.11$ & $43.2$ $\pm 13.37$ & $0.163^{+ 0.067}_{-0.059}$  & $25.62 \pm 0.06$ & $0.47 \pm 0.06$\\ 
$1529106057$ & $220.3720$ & $-0.1072$ & $6.0464$ & $43.27\pm0.01$  &  $262.15 \,\pm 21.19$ & $18.08$ & $38.92$ $\pm 7.83$ & $0.147^{+ 0.043}_{-0.038}$ & $25.79 \pm 0.04$ & $0.63 \pm 0.06$ \\ 
%\midrule
%1206233254 & 6.150 & 6.607 & 6.1243 & 42.20  &  $211.80 \,(\pm 21.19)$ & $19.74$ ($\pm 6.15$) & $0.27$ \\ 
%\midrule
$2310222098$ & $348.3013$ & $-34.0200$ & $6.1485$ & $43.40\pm 0.01$  &  $329.61 \,\pm 20.75$ & $11.07$ & $65.81$ $\pm 9.58$ & $0.241^{+ 0.057}_{-0.052}$ & $-$ & $0.59 \pm 0.05$\\ 
%\midrule
$1528094186$ & $219.5400$ & $-1.1004$ & $6.1662$ & $42.78 \pm 0.04$  &  $230.72 \,\pm 26.26$ & $15.77$ & $12.65$ $\pm 4.23$ & $0.047^{+ 0.020}_{-0.018}$ & $25.83 \pm 0.18$ & $0.57 \pm 0.05$\\ 
%\midrule
$1505103163$ & $214.6632$ & $-1.7198$ & $6.2481$ & $42.72 \pm 0.05$  &  $221.43 \,\pm 47.02$ & $12.26$ & $13.73$ $\pm 2.55$ & $0.051^{+ 0.014}_{-0.012}$ & $25.01 \pm 0.01$ & $0.57 \pm 0.03$\\ 
%\midrule
%1523213176 & 219.5400 & -1.1004 & 6.4030 & 42.77  &  $170.12 \,\pm 17.15$ & $11.23$ $\pm 7.27$ & $0.05 \, \pm 0.01$ \\ 
%\midrule
$1530068179$ & $222.1510$ & $2.9404$ & $6.4202$ & $42.80\pm 0.04$ & $247.26 \,\pm 20.11$ & $4.97$ & $37.43$ $\pm 12.97$ & $0.141^{+ 0.063}_{-0.056}$ & $< 26.06$ & $0.42 \pm 0.06$\\ 
%\midrule
$1528263095$ & $220.2286$ & $-1.6538$ & $6.5540$ & $42.57 \pm 0.06$ &  $163.44 \, \pm 21.19$ & $6.53$ & $16.83$ $\pm 4.77$ & $0.063^{+ 0.024}_{-0.021}$ & $< 26.03$ & $0.45 \pm 0.04$\\
%\midrule
$1522272275$ & $219.0635$ & $0.8046$ & $6.6073$ & $43.32\pm 0.02$ &  $325.20 \, \pm 20.33$ & $32.84$ & $16.97$ $\pm 3.57$ & $0.063^{+ 0.019}_{-0.017}$ & $26.29 \pm 0.08$ & $0.80 \pm 0.07$
\\ 
\bottomrule
\end{tabular}
\end{threeparttable}
\end{table*}

\subsection{HSC photometry}
We use optical photometric data for $17$ MAGPI LAEs of our sample that are covered in the broad-band filters ($g$, $r$, $i$, $z$ and $y$) of Subaru HSC Wide layer. HSC Strategic Program \citep[see][]{Aihara18} is a wide-field optical imaging survey on the $8.2$ meter Subaru Telescope. The HSC-Wide layer data cover about $300$ $\mathrm{deg}^2$ in all five broad-band filters to the nominal survey exposure ($10$ min in $g$ and $r$ bands and $20$ min in $i$, $z$, and $y$ bands; see \citealt{Aihara19}) with a median seeing $0.6''$ in the $i$-band.

Using HSC command-line SQL (Structured Query Language) tool \footnote{\href{https://hsc-release.mtk.nao.ac.jp/datasearch}{https://hsc-release.mtk.nao.ac.jp/datasearch}}, we retrieve the Wide layer photometric data from data-release $2$ \citep{Aihara19}. We extract fluxes and corresponding $1 \sigma$ flux uncertainties and limiting $2\sigma$ fluxes (for non-detections) using $2''$ aperture diameter. These fluxes and corresponding uncertainties for $i$, $z$ and $y$ bands are presented in Table \ref{t1}. 

\section{Data analyses}\label{sec:analyses}
For our sample of $22$ LAEs, 1D spectra are extracted using an aperture of $2''$ radius. MUSE 1D spectra reveal the detection of  \lya\ emissions where spectroscopic redshifts are determined based on the peak of the \lya\ line. No other associated emission lines are found in the MUSE data for these $22$ LAEs, restricting the measurement of systemic redshift. Below we estimate UV magnitudes and discuss the procedures of analysing \lya\ spectra to study properties of these LAEs. 

\subsection{Estimating UV magnitudes}\label{3.1}
Out of the $22$ sources, $11$ LAEs are detected in HSC broad-band filters. HSC does not cover GAMA$23$ fields and hence we do not have photometric data for four sources in this field. We also do not have photometric data for MAGPI$1207184066$ as it is located at the edge of the HSC field. For rest of the six sources, which are not detected in HSC, we calculate a $2\, \sigma$ limit on the magnitudes (see Table \ref{t1}). As we do not have photometry beyond $y$ band,  no meaningful constraints can be obtained on the UV
slope ($\beta$). Therefore, we assume a flat UV slope $\beta = -2$, which is the typical value that most of the high-redshift galaxies have \citep[see,][]{Dunlop13, Bouwens14, Matthee17}. We convert the $y$ band magnitudes to the rest-frame absolute UV magnitudes ($\mathrm{M}_{\mathrm{UV}}$) at $1500$ \AA. We also note that, for $z>6$ sources, y-band magnitude is not exactly measuring the flux at $1500$ \AA. Still, it provides a decent approximation of the UV magnitudes. The estimated UV magnitudes of our sources lie in the range $-19.74 \lesssim \mathrm{M}_{\mathrm{UV}} \lesssim -23.27$, which are given in Table \ref{t1}.

\subsection{Line profile fitting}\label{3.2}
% The \lya\ line in MR5 shows a simple red-skewed profile. However, the profiles of MR6 and MR7 are more complex, where MR6 exhibits a skewed profile with an extended red tail, whereas MR7 displays a wide main profile with a narrow red wing with $ 3 \sigma$ detection. 

We use \texttt{pyplatefit} \footnote{\href{https://github.com/musevlt/pyplatefit}{https://github.com/musevlt/pyplatefit}}, a python module \citep{Bacon23} to fit a local continuum and obtain a continuum subtracted spectrum. \texttt{pyplatefit} performs a continuum fit around the observed \lya\ line in a spectral window of $\pm 50$ \AA\ centred on the \lya\ line, using a simple stellar population model \citep{Bruzual2003}. It then subtracts the continuum and returns a continuum-subtracted spectrum. Next we fit the continuum subtracted \lya\ profiles using an asymmetric Gaussian which has been found to provide an extremely
good representation of the \lya\ line \citep[see][]{Shibuya14, Herenz17, Claeyssens19, Songaila24}:
\begin{equation}
    F \, (\lambda) = f_{\mathrm{max}} \, \mathrm{exp} \left(-\, \frac{\Delta  v^{\, 2}}{2(a_{\mathrm{asym}}\,(\Delta v)+w)^2}\right)
\end{equation}
where $f_{\mathrm{max}}$ is the flux normalization (amplitude), $\Delta v$ is the velocity shift (in $\mathrm{km}\, \mathrm{s}^{-1}$) relative to the peak velocity, $a_{\mathrm{asym}}$ determines the asymmetry or skewness of the line and $w$ (in $\mathrm{km}\, \mathrm{s}^{-1}$) controls the line width (FWHM). A positive asymmetry value suggests that the \lya\ line has a red wing, which is usually seen in most of the single-peaked \lya\ lines \citep{Kerutt22, Songaila22}. The line width can be obtained in terms of fitting parameters as follows \citep[see also,][]{Claeyssens19, Songaila24}: 
\begin{equation}
    \mathrm{FWHM}\, (\mathrm{km}\, \mathrm{s}^{-1}) = \frac{2\sqrt{2\,\mathrm{ln}(2)} \, w}{(1 - 2\,\mathrm{ln}(2) \, a^{\,2}_{\mathrm{asym}} )}
\end{equation}
 The corresponding error in the
FWHM is almost fully dominated by the error in $w$, with only a few percent contribution from the asymmetry term. The fits on the continuum-subtracted spectra along with the values of the free parameters and corresponding $1 \sigma$ errors are shown in \ref{app1}. 

\subsection{Spectroscopic measurements}\label{3.3}

\begin{figure}[h!]
    \centering
    \includegraphics[width=0.95\textwidth]{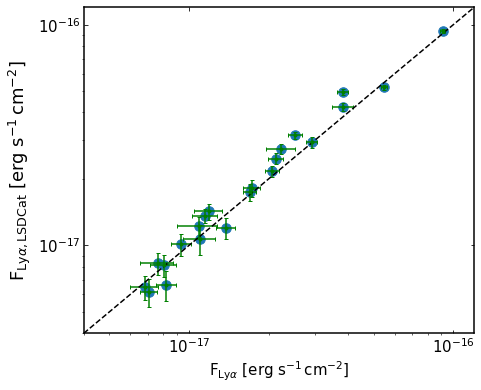}
    \caption{\lya\ line fluxes obtained from MUSE 1D spectra (extracted using $2''$ radius aperture) are compared against \texttt{LSDCat} $3\, R_{\mathrm{KRON}}$ fluxes. The one-to-one relation is shown as the dashed line. }
    \label{f1}
\end{figure}

Using a single asymmetric Gaussian fit to the \lya\ profile extracted using $2''$ radius aperture, as mentioned above, we measure \lya\ line fluxes ($F_{\mathrm{Ly}\alpha}$) for our LAEs. In Fig. \ref{f1}, we compare them with the $3\, R_{\mathrm{KRON}}$ fluxes obtained using \texttt{LSDCat} (flux extracted using the aperture of radius $3\times  R_{\mathrm{KRON}}$ that contains $> 95$\% of the total line flux; \citep[see,][]{Graham05, Herenz17}. The values of $3\, R_{\mathrm{KRON}}$ radii in which fluxes were extracted are ranging from $1.6''$ to $2.2''$. Since both flux values match well, we conclude that our line flux measurements are accurate.  The observed \lya\ luminosities ($L_{\mathrm{Ly}\alpha}$) are then calculated from fluxes as $L_{\mathrm{Ly}\alpha} = 4 \pi \, F_{\mathrm{Ly}\alpha} D^{\,2}_{\mathrm{L}} $, where $D_{\mathrm{L}}$ is the luminosity distance, calculated using the cosmological parameters that we assume in \S \ref{sec:intro}. 

We measure the \lya\ $\mathrm{EW}_{0}$ to investigate the strength of the \lya\ line.  
%In order to do that, we first fit local continuum around the \lya\ line in a spectral window of $\pm 50$ \AA\ centred on the \lya\ line using \texttt{pyplatefit} PYTHON package \citep{Bacon23}. 
In order to do that, we need measurements of UV continuum flux density at the \lya\ wavelength. However, due to lack of UV-slope measurement (see \S \ref{3.1}), it is not possible to determine continuum flux density from the photometric data. Further, \cite{Hashimoto17} caution that fixing the value of UV-slope $\beta$ at high-redshift can lead to an underestimation of $\mathrm{EW}_{0}$ due to the redshift evolution of $\beta$. Therefore, we obtain the observed median UV-continuum flux density ($f^{\mathrm{cont}}_{\mathrm{Ly}\alpha}$) from the continuum fit performed using \texttt{pyplatefit} (see \S \ref{3.2}). We then divide the total \lya\ line flux ($ \mathrm{F}_{\mathrm{Ly}\alpha} $) by $f^{\mathrm{cont}}_{\mathrm{Ly}\alpha}$ to determine observed \lya\ equivalent width (EW). The rest-frame equivalent width is then given as $\mathrm{EW}_{0} = \mathrm{EW} / (1 + z)$. We note that for some LAEs, the continuum in the MUSE data is too faint, and thus we can only derive lower limits for $\mathrm{EW}_{0}$ for them. As \ha\ emission is not covered within the MUSE spectral range, \lya\ escape fraction can not be directly measured. A strong correlation between $f^{\mathrm{Ly}\alpha}_{\mathrm{esc}}$ and $\mathrm{EW}_{0}$ has been found in both low- and high-redshift LAEs \citep[see,][]{Matthee17, Yang17, Sobral19, Begley24}. In \cite{Begley24}, a sample of $152$
star-forming galaxies with $z \sim 4 - 5$ is used to obtain a linear dependence between $f^{\mathrm{Ly}\alpha}_{\mathrm{esc}}$ and $\mathrm{EW}_{0}$. We estimate $f^{\mathrm{Ly}\alpha}_{\mathrm{esc}}$ using this best-fit relation. We tabulate the spectroscopic properties of these LAEs in Table\,\ref{t2}. 

\cite{Begley24} discuss that their $f^{\mathrm{Ly}\alpha}_{\mathrm{esc}}$ - $\mathrm{EW}_{0}$ relation agrees well with the relation derived for low-redshift LAEs \citep{Sobral19}. The observed scatter in this relation are found to be well-consistent with that observed in both low and high-$z$ LAE samples \citep[see,][]{Pucha22, Roy23}. A similar positive correlation has also been observed in LAEs in the reionization era \citep{Saxena24, Tang24b}. However, the slope, normalization and scatter of the observed $f^{\mathrm{Ly}\alpha}_{\mathrm{esc}}$ - $\mathrm{EW}_{0}$ relationship are likely influenced by dust attenuation, differential dust geometry and $\xi_{\mathrm{ion}}$ \citep{Matthee17a, Harikane18, Shivaei18, Sobral19}. Additionally, for a given $\mathrm{EW}_{0}$ value, intrinsic scatter in $f^{\mathrm{Ly}\alpha}_{\mathrm{esc}}$ is expected due to variations in stellar populations and dust/gas properties \citep{Begley24}. We conclude that, despite the complexity of factors, $f^{\mathrm{Ly}\alpha}_{\mathrm{esc}}$ can be predicted within $\sim 0.5$ dex of actual values from the $f^{\mathrm{Ly}\alpha}_{\mathrm{esc}}$ - $\mathrm{EW}_{0}$ relation of \cite{Begley24} using only the equivalent width information.

\begin{figure*}
\centering
    {\includegraphics[width=8.5cm,height=6.5cm]{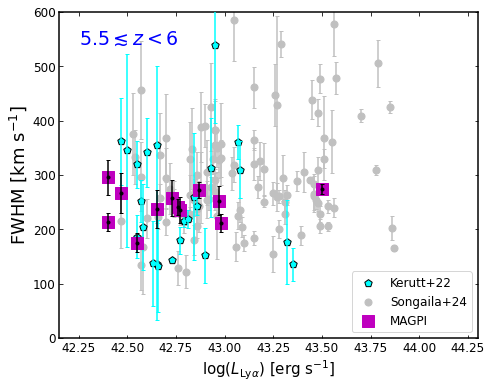}}
    {\includegraphics[width=8.5cm,height=6.5cm,clip]{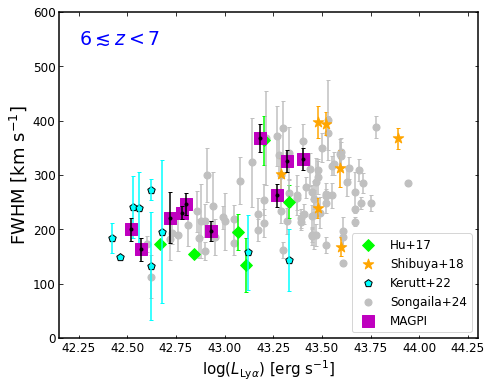}}
\caption{Evolution of \lya\ line width is shown as function of \lya\ luminosity at $5.5 \lesssim z < 6$ (left panel) and at $6 \lesssim z \lesssim 7$ (right panel). MAGPI LAEs (this work) are shown as purple squares. We also include data from MUSE DEEP and MUSE WIDE surveys \citep[blue pentagons;][]{Kerutt22}, \citep[grey circles;][]{Songaila24}, LAGER survey \citep[green diamonds;][]{Hu17} and \citep[orange stars;][]{Shibuya18}.}
\label{fig2}
\end{figure*}

%\begin{figure*}[hbt!]
%\centering
%    {\includegraphics[width=8.5cm,height=6.5cm]{Images/XivsZ.png}}
%    {\includegraphics[width=8.5cm,height=6.5cm,clip]{Images/RBvsFWHM.png}}
%\caption{{\it Left panel:} Evolution of ionizing photon production efficiency as a function of redshift. The yellow stripe shows the canonical values from \citep{Kuhlen12, Robertson13}. The blue dotted line presents the median $\xi_{\mathrm{ion}} = 25.25 \, \mathrm{Hz}\, \mathrm{erg}^{-1}$ for our MAGPI LAEs. The green and orange dashed lines are the median $\xi_{\mathrm{ion}}$ values for JADES LAEs \citep{Saxena24} and JEMS + MUSE LAEs \citep{Simmonds23} respectively. {\it Right panel:} Evolution of the bubble radius as a function of \lya\ line width. LAEs at $5.5 \lesssim z < 6$ are highlighted in green squares whereas LAEs at $z > 6$ are shown in purple squares. The linear fit to the data is shown as black dashed line. The significance of the corresponding fit is given on the top left corner.}
%\label{fig4}
%\end{figure*}

%By default, this template uses \texttt{biblatex} and adopts the Chicago referencing style. If you are using this template on Overleaf, Overleaf's build tool will automatically run \texttt{pdflatex} and \texttt{biber}. If you are compiling this template on your own local \LaTeX{} installation, please execute the following commands:
%\begin{enumerate}
%    \item \verb|pdflatex sample|
%    \item \verb|biber sample|
%    \item \verb|pdflatex sample|
%    \item \verb|pdflatex sample|
%\end{enumerate}

\section{Results and discussions}\label{sec:results}

We present photometric and spectroscopic properties (see Table \,\ref{t1} and Table \,\ref{t2} respectively) of 22 LAEs at $5.5 \lesssim z \lesssim 6.6$. These sources have \lya\ luminosities $\mathrm{log} (L_{\mathrm{Ly}\alpha} \, [\mathrm{erg} \, \mathrm{s}^{-1}]) = 42.4 - 43.5 $ and \lya\ rest-frame equivalent widths $\mathrm{EW}_{0} \simeq 10 - 72$ \AA. In the following subsections, we explore the relation between the \lya\ line width and \lya\ luminosity and investigate this relation in the context of the size of the ionized bubbles created by each LAE at the end of the reionization era. 

\subsection{Evolution of the \lya\ line width}\label{4.1}

We fit the \lya\ lines with an asymmetric/skewed Gaussian. This provides excellent representation of all the \lya\ profiles. All of the asymmetry parameters are positive, corresponding to red-skewed profiles. The asymmetry of all the lines also confirm that these are \lya\ emission from star-forming galaxies rather than other emission lines from low-redshift interlopers. In Fig. \ref{fig2}, we present the evolution of \lya\ line widths with \lya\ luminosities across two redshifts ranges. For MAGPI LAEs, we find that at $z < 6$ (see Fig \ref{fig2}, left panel), FWHM distribution is almost uniform, with no significant evolution with increasing luminosities. In contrast, $z > 6$ LAEs show a strong evolution of increasing line width with increasing luminosities (see Fig. \ref{fig2}, right panel). For comparison, we also include data from MUSE DEEP and MUSE WIDE surveys \citep{Kerutt22}, bright LAEs from \cite{Shibuya18}, $z \sim 6.9 $ LAEs from LAGER survey \citep{Hu17} and recent large sample of LAEs from \cite{Songaila24}, which allows a better understanding of the evolution of line widths across a wide redshift range ($5.5 \lesssim z \lesssim 7$) at the edge of reionization. We note large uncertainties in luminosities and FWHMs in the data of \cite{Kerutt22}. Apart from that, we find our result is well-consistent with the others. We note that, at $z > 6$, line widths of MAGPI LAEs are narrower (FWHM $\lesssim 250 \, \mathrm{km}\, \mathrm{s}^{-1}$) at luminosities $\mathrm{log}(L_{\mathrm{Ly}\alpha}\, [\mathrm{erg}\, \mathrm{s}^{-1}]) < 43 $, as compared to LAEs at $\mathrm{log}(L_{\mathrm{Ly}\alpha}\, [\mathrm{erg}\, \mathrm{s}^{-1}]) > 43 $, where we find broader line widths (i.e. FWHM $> 260 \, \mathrm{km}\, \mathrm{s}^{-1}$). The visibility of \lya\ emission from high-redshift star-forming galaxies is primarily determined by the star formation rate and ionizing photon budget, which influence the size of ionized bubbles around them that allow \lya\ photons to travel unattenuated along the line of sight \citep{MR06}. Despite the fact that increasingly neutral IGM potentially scatters most of the \lya\ photons and make the line narrower, the broadening of the line at $z > 6$ suggests that the high-luminosity LAEs may preferentially lie in more highly ionized regions than the lower luminosity LAEs \citep{Matthee17, Songaila24}, protecting themselves from the scattering effects of the IGM. In the next section, we discuss how LAEs themselves can produce ionized bubbles around them. 

%We try to improve the understanding of the evolution of \lya\ line width with luminosity across a wide redshift range ($ 5.5 \lesssim z \lesssim 7$) by expanding the sample of \cite{Songaila22}, where this study was limited to LAEs at $z = 5.7$ and at $z = 6.6$ only. 
%We find an increase in the slope of the linear fit on the \lya\ line width (FWHM) vs luminosity plot for LAEs at $z > 6$, as shown in Fig.\,\ref{fig3}. We find a strong correlation at $6 < z < 7$ between line width and luminosity with a very high significance ($\rho = 0.66, p = 3.96 \times 10^{-8}$). This correlation is less significant at $5.5 \lesssim z < 6$ ($\rho = 0.31, p = 1.51 \times 10^{-3}$).  This result is consistent with the previous findings \citep[see][]{Matthee17, Songaila22}. 

\subsection{Ionizing photon escape and size of ionized region}\label{4.2}

The absorption of ionizing photons by the interstellar medium within galaxies leads to \lya\ emissions through recombination processes, while photons that escape contribute to cosmic reionization. 
The size of the ionized region (\hii\ bubble) around an ionizing source can be estimated by solving the evolution equation of the ionizing front \citep[e.g.][]{Cen2000, Yajima18} : \\
\begin{equation}
    \frac{\mathrm{d}R^{3}_{\mathrm{B}}}{\mathrm{d}t} = 3\, H (z)\, R^{3}_{\mathrm{B}} + \frac{3\, Q_{\mathrm{ion}}\,f^{\mathrm{LyC}}_{\mathrm{esc}}}{4\pi\,n_{\mathrm{H}}(z)} - C_{\mathrm{HII}}\,n_{\mathrm{H}}(z)\,\alpha_{\mathrm{rec}}\,R^{3}_{\mathrm{B}}
\end{equation}
where $R_{\mathrm{B}}$ is the proper physical radius of ionized bubble in physical Mpc (pMpc), $H (z)$ is the Hubble parameter, $Q_{\mathrm{ion}}$ is the intrinsic production rate of ionizing photons (in $\mathrm{s}^{-1}$), $C_{\mathrm{HII}}$ is the clumping factor of ionized hydrogen and $\alpha_{\mathrm{rec}}$ is the temperature-dependant total recombination rate coefficient under Case B approximation (T = $10^{4}$ K, $n_{e} = 350 \mathrm{cm}^{-3}$). The mean hydrogen density of the IGM ($n_{\mathrm{H}}$) scales with redshift as follows: $n_{\mathrm{H}} \approx 8.5 \times 10^{-5} \, \left(\frac{1 + z}{8}\right)^{3} \, \mathrm{cm}^{-3}$ \citep[see,][]{MG20, Meyer21}. 

% A clear correlation between $f^{\mathrm{Ly}\alpha}_{\mathrm{esc}}$ and $f^{\mathrm{LyC}}_{\mathrm{esc}}$ has been found in confirmed LyC leakers as well as in simulations \citep[e.g.][]{Verhamme17, Chisholm18, Kimm19, Flury22, Begley24}. We estimate $f^{\mathrm{LyC}}_{\mathrm{esc}}$ based on the best-fitting relation between $f^{\mathrm{Ly}\alpha}_{\mathrm{esc}}$ and $f^{\mathrm{LyC}}_{\mathrm{esc}}$ \citep[see, Fig. 7 of][]{Begley24}. 
We can define ionizing photon production rate ($Q_{\mathrm{ion}}$) in terms of direct \lya\ observables \citep[see,][]{Matthee17, Yajima18, Sobral19, Matthee22}:
\begin{equation}
    Q_{\mathrm{ion}} \, [s^{-1}] = \frac{L_{\mathrm{Ly}\alpha}}{c_{\mathrm
    {H}\alpha}\, \left(1 -f^{\mathrm{LyC}}_{\mathrm{esc}}\right)\, \times \left(8.7\, f^{\mathrm{Ly}\alpha}_{\mathrm{esc}}\right)}
\end{equation}
where $c_{\mathrm{H}\alpha} = 1.36 \times 10^{-12} $ erg \citep[under case B recombination;][]{K98, Schaerer2003}. Here, we assume negligible nebular attenuation \citep{Naidu22}. A detailed discussion on the role of dust in the context of \lya\ and LyC escape can be found in \cite{Kakiichi21}. In \S \ref{3.3}, we estimated \lya\ escape fractions (within $\simeq 0.5$ dex) using its linear dependence on $\mathrm{EW}_0$. However, this relation can be influenced by dust attenuation and $\xi_{\mathrm{ion}}$, as discussed in \S \ref{3.3}. A prominent degeneracy between dust extinction and $\xi_{\mathrm{ion}}$ has been observed in \cite{Sobral19}, where higher dust extinction allows for a lower $\xi_{\mathrm{ion}}$ and vice versa.  Current data restrict us from accurately estimating both $f^{\mathrm{Ly}\alpha}_{\mathrm{esc}}$ and $\xi_{\mathrm{ion}}$ as it requires dust-corrected \ha\ luminosity. Direct observations of Balmer decrements and high-excitation UV lines are necessary to further validate and confirm our results. 

A correlation between \lya\ and LyC escape fractions has been investigated in several observational studies and hydrodynamical simulations \citep{Verhamme17, Chisholm18, Flury22, Maji22}. \cite{Begley24} find a linear dependence between $f^{\mathrm{LyC}}_{\mathrm{esc}}$ and $f^{\mathrm{Ly}\alpha}_{\mathrm{esc}}$. Using this relation, we estimate that our sources have $f^{\mathrm{LyC}}_{\mathrm{esc}} = 0.5 - 4$ \%. Hydrodynamical and radiative transfer simulations
are used to calibrate the same relation \citep{Maji22}, using which we get $f^{\mathrm{LyC}}_{\mathrm{esc}} < 3$ \% for our sources. However, this correlation between \lya\ and LyC escape does not seem to
work well for LyC leakers at higher redshifts \citep[see,][]{Kerutt24}. Kinematics and ISM properties of high-redshift leakers are more complex \citep{Guaita15} as compared to low-redshift leakers. A spatial offset between \lya\ and LyC emission has been observed in high-redshift leakers \citep{Kerutt24, Gupta24}, which indicates that the two are escaping from different locations in the galaxy. For instance, high-redshift leakers prefer asymmetric escape ( indicating scattering or expanding gas) rather than through optically thin ionized channel, which seems to be the case for low-redshift leakers \citep{Kerutt24}. 

We do not expect LAEs in our sample to be strong LyC leakers. Hence, we simply assume a fiducial value of LyC escape $f^{\mathrm{LyC}}_{\mathrm{esc}} = 5$ \% for our sources throughout the analysis. The ionizing photon production efficiency is defined as the ratio of $Q_{\mathrm{ion}}$ and intrinsic UV luminosity density ($L_{\mathrm{UV},\nu}$) \citep{Matthee17} :
\begin{equation}
    \xi_{\mathrm{ion}} \, [\mathrm{Hz}\, \mathrm{erg}^{-1}] = \frac{Q_{\mathrm{ion}}}{L_{\mathrm{UV},\nu}}
\end{equation}
 We obtain $L_{\mathrm{UV},\nu}$ from $\mathrm{M}_{\mathrm{UV}}$, assuming negligible dust attenuation. We refer to \cite{Bouwens16} for a discussion on the impact of dust attenuation on $\xi_{\mathrm{ion}}$. Our measured $\mathrm{M}_{\mathrm{UV}}$ values (see Table \ref{t1}) indicate that these LAEs are UV-bright ( $ -19.74 \leq \mathrm{M}_{\mathrm{UV}} \leq -23.27$). The estimated $\xi_{\mathrm{ion}}$ values are presented in Table \ref{t2}. For the sources, which are detected in HSC, we find an average ionizing photon production efficiency $\mathrm{log}(\xi_{\mathrm{ion}}\, [\mathrm{Hz}\, \mathrm{erg}^{-1}]) = 25.51$ at $z < 6$, while $\mathrm{log}(\xi_{\mathrm{ion}}\, [\mathrm{Hz}\, \mathrm{erg}^{-1}]) = 25.74$ at $z > 6$. The slight evolution of $\xi_{\mathrm{ion}}$ with redshift is consistent with the previous studies \citep[eg.][]{Bouwens14, Endsley2, Simmonds23, Tang23, Saxena24}. This aligns with the idea that younger galaxies may achieve higher ionizing photon production efficiencies. However, the evolution is very mild which could suggest that the production and escape of ionizing photons are governed by physical processes operating on much shorter timescales, such as intense star formation or supernova activity, which do not show a strong trend with redshift \citep{Saxena24}. Further, we also note that some of the low-equivalent width (therefore, low $f^{\mathrm{Ly}\alpha}_{\mathrm{esc}}$) LAEs ($\mathrm{EW}_0 < 20$ \AA; i.e, MAGPI$1503111271$, MAGPI$1204117107$, MAGPI$1528094186$, and MAGPI$1522272275$) show higher $\xi_{\mathrm{ion}}$ (see Table \ref{t2}). Such a tentative anti-correlation between $f^{\mathrm{Ly}\alpha}_{\mathrm{esc}}$ and $\xi_{\mathrm{ion}}$ has been reported in a large sample of LAEs from JEMS and MUSE \citep[see][]{Simmonds23}. This could indicate a time delay between production and escape of ionizing photons in these galaxies \citep{Katz20}, which again may be linked to supernova activity. 

\begin{figure}[h!]
    \centering
    \includegraphics[width=1.0\textwidth]{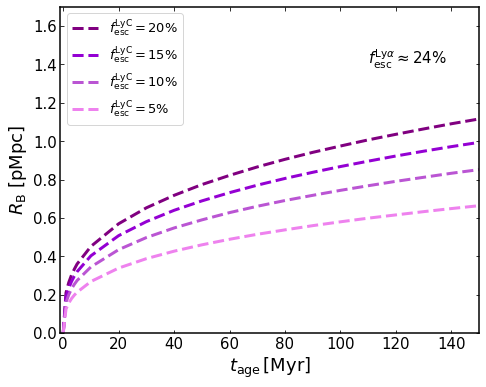}
    \caption{Radius of spherical bubble ionized by MAGPI$2310222098$ as a function of the age parameter ($t_{\mathrm{age}}$), for different values of the LyC escape fraction ($f^{\mathrm{LyC}}_{\mathrm{esc}} = 5$ \%, $10$ \%, $15$ \% and $20$ \%). Here we assume a fixed \lya\ escape fraction $f^{\mathrm{Ly}\alpha}_{\mathrm{esc}} = 24$\%, as estimated (see Table \ref{t2}).}
    \label{f3}
\end{figure}

For constant \lya\ luminosity and $f^{\mathrm{LyC}}_{\mathrm{esc}}$, hence for a constant production rate $Q_{\mathrm{ion}}$, Eq. (3) can be solved analytically to obtain an expression of bubble sizes that can be produced by the LAE on its own. For instance, neglecting the accelerated expansion due to Hubble flow and for luminous sources at $z \lesssim 8$, when the recombination rate is relatively low, second term of Eq. (3) dominates and we get an expression for bubble radius \citep{Cen2000}:
\begin{equation}
    R_{\mathrm{B}} \approx \left(\frac{3\, Q_{\mathrm{ion}}\,f^{\mathrm{LyC}}_{\mathrm{esc}}\, t_{\mathrm{age}} }{4\pi\,n_{\mathrm{H}}(z)}\right)^{1/3}
\end{equation}
where $t_{\mathrm{age}}$ is the time since the ionizing source has switched on. For these estimates, we assume a fiducial value of $t_{\mathrm{age}} = 100$ Myr \citep[a reasonable amount of time for $f^{\mathrm{LyC}}_{\mathrm{esc}} = 5$ \%; see][]{Witstok24, Whitler24} for all the LAEs. Recent studies note that the inferred bubble size does not strongly depend on small deviations in LyC escape fraction and age \citep[see,][]{Witstok24, Torregrosa24}. For one luminous LAE (ID: MAGPI$2310222098$), we study the evolution of bubble size as a function of $t_{\mathrm{age}}$ for different values of the LyC escape fraction (see, Fig. \ref{f3}). For a constant \lya\ escape fraction ($f^{\mathrm{Ly}\alpha}_{\mathrm{esc}} \approx 24$ \%), we find that this LAE is capable of ionizing a bubble of radius $R_{\mathrm{B}} \sim 0.59$ pMpc in $t_{\mathrm{age}} = 100$ Myr for a LyC escape of $f^{\mathrm{LyC}}_{\mathrm{esc}} = 5$ \% while it takes $t_{\mathrm{age}} = 50\, (30)$ Myr to ionize the same bubble size, when $f^{\mathrm{LyC}}_{\mathrm{esc}} = 10\, (15)$ \%. We also note that $t_{\mathrm{age}}$ depends on the actual time duration of constant star-formation within the galaxy, which is fundamentally bounded to $\lesssim 200$ Myr at this epoch \citep[see,][]{Tacchella18, Whitler23}.

\begin{figure}[h!]
    \centering
    \includegraphics[width=1.0\textwidth]{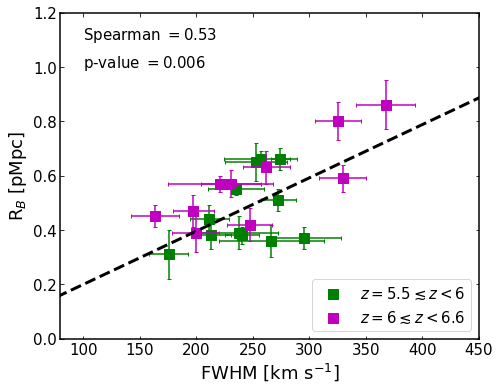}
    \caption{Evolution of the bubble radius as a function of \lya\ line width for our 22 MAGPI LAEs. LAEs at $5.5 \lesssim z < 6$ are highlighted in green squares whereas LAEs at $z > 6$ are shown in purple squares. Significance of the plot (Spearman correlation coefficient and p-value) are shown in the top left corner.}
    \label{f4}
\end{figure}

%During the recombination time scale, ionizing front can travel up to the Stromgren radius, $R_{\mathrm{S}} = \left(\frac{3\, Q_{\mathrm{ion}}\,f^{\mathrm{LyC}}_{\mathrm{esc}}}{4\pi\, \alpha_{\mathrm{rec}} \, n^{2}_{\mathrm{H}}(z)}\right)^{1/3}$.

Based on the assumptions mentioned above ($t_{\mathrm{age}} = 100$ Myr and $f^{\mathrm{LyC}}_{\mathrm{esc}} = 5$ \%), we find that each of our LAEs reside in ionized bubbles of radii $R_\mathrm{B} = 0.31 - 0.86$ pMpc (see, Table \,\ref{t2}). From Eq. (4) and Eq. (6), we note that bubble radius scales with the \lya\ luminosity as $R_\mathrm{B} \propto L^{1/3}_{\mathrm{Ly}\alpha}$. Modeling of \lya\ emission and transmission during the EoR also predicts that \lya\ luminosity increases with $R_{\mathrm{B}}$ due to higher IGM transmission for large ionized bubbles \citep[see,][]{Yajima18}. We find a linear correlation between bubble radii and \lya\ line widths (see Fig. \,\ref{f4}) with high statistical significance (Spearman correlation coefficient $= 0.53$ and p-value $= 0.006$). This trend is more prominent in LAEs at $z > 6$, which is expected given the correlation between \lya\ luminosity and line width at $z > 6$, as discussed in the previous section. This again supports the fact that broadening of \lya\ lines at $z > 6$ is due to large ionized bubble which are created around them, allowing most of the \lya\ photons to come out of that the galaxy without suffering from much scattering in the IGM \citep{Songaila22, Songaila24}.

At high redshifts, \lya\ lines are usually redshifted with respect to systemic velocity due to strong outflows, which facilitates the transmission of \lya\ photons through IGM \citep{Dijkstra11}. The velocity offset of the red peak has been used to place lower limits on the bubble sizes required for IGM transmission \citep{MG20, Witstok24}. Considering a patchy reionization scenario, where a galaxy is typically surrounded by a completely neutral IGM, it is found that \lya\ can be detected at a high velocity offset of $\gtrsim 300 \, \mathrm{km}\, \mathrm{s}^{-1}$ when the galaxy is situated in a ionized region of radius $R_{\mathrm{B}} \gtrsim 0.1$ pMpc \citep[see][]{MG20, Umeda23}. A Bayesian approach of modeling intrinsic \lya\ profiles has predicted the size of ionized regions to be $R_{\mathrm{B}} = 0.5 - 2.5$ pMpc at $z > 6$ \citep{Hayes23}. \cite{Witstok24} find that LAEs with relatively low \lya\ velocity offset ($\lesssim 300 \, \mathrm{km}\, \mathrm{s}^{-1}$) with moderately high escape fraction can represent ionized regions of sizes $R_{\mathrm{B}} = 0.1 - 1$ pMpc. Further, several attempts have been made to put constraints on the size of ionized regions that allow blue peak of \lya\ to be detected at $z > 6$, where it is found that the blue peak can be detected if the source galaxy resides in a highly ionized region ($x_{\mathrm{HI}} > 10^{-5}$) of radius $R_{\mathrm{B}} \gtrsim 0.5 $ pMpc \citep[see,][]{MG20, Torregrosa24}. We note that, while some of our LAEs show larger bubble sizes, we can not always anticipate seeing a blue-peak emission from them, as these photons might be heavily absorbed by neutral hydrogen in the circumgalactic medium \citep[CGM;][]{Henry15, Gazagnes20, Endsley22}. 

\begin{figure*}[ht!]
\centering
    {\includegraphics[width=8.2cm,height=6.3cm]{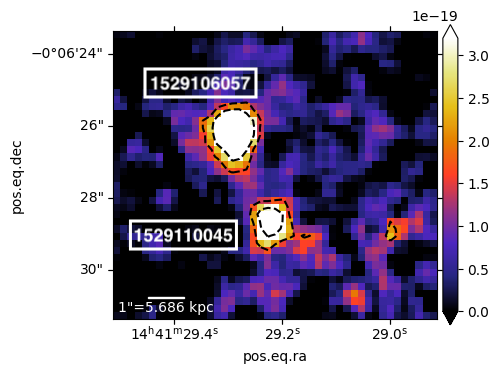}}
    {\includegraphics[width=7cm,height=7cm,clip]{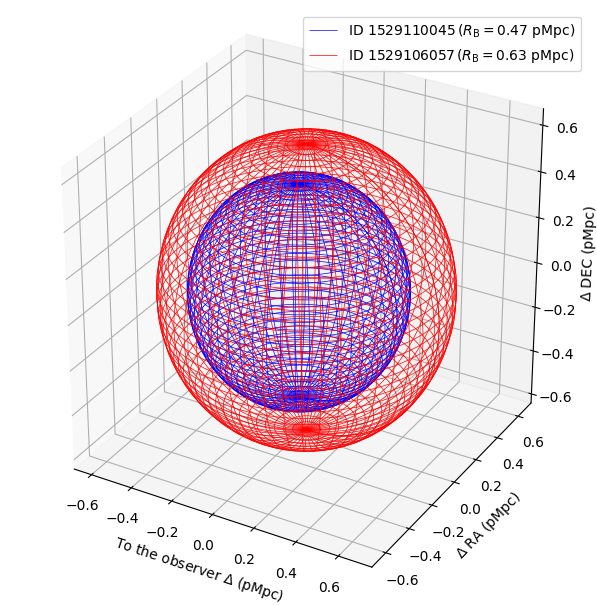}}
\caption{{\it Left panel:} Synthetic MUSE \lya\ narrow band (NB) image collapsed for wavelengths within $8550$ \AA\ to $8590$ \AA, showing two LAEs with MAGPI IDs $1529106057$ and $1529110045$ at $z = 6.046$, separated by a physical distance of $15.92$ kpc. Contours are shown as black dashed lines at the $2$ and $4 \sigma$ significance levels. {\it Right panel:} Three-dimensional visualization of the size of ionized bubbles created by them along the line of sight. Bubbles are shown in wire-frame spheres centred on each LAE. The centre of the LAE MAGPI$1529110045$ is at the origin of this 3D plot. }
\label{f5}
\end{figure*}

\lya\ radiative transfer modeling suggests strong correlation between \lya\ velocity offset and $N_{\mathrm{HI}}$ \citep{Verhamme15}. As $N_{\mathrm{HI}}$ increases the peak velocity shifts farther from the systemic velocity. Further, shell model predicts that for a low $N_{\mathrm{HI}}$ and high outflow velocity, most of the \lya\ photons can directly escape at the line centre frequency \citep{Yajima18}. In addition, line width becomes narrower when $N_{\mathrm{HI}}$ decreases. This is expected as low $N_{\mathrm{HI}}$ shifts the peak to shorter wavelengths, where \lya\ flux is significantly reduced due to IGM scattering. While IGM radiative transfer establishes a typical correlation between the line width and velocity offset of \lya\ red peaks \citep[e.g.][]{Neufeld90, Verhamme18, Li22}, this trend is expected to change at high redshifts due to IGM absorption. Ignoring such complicated radiative transfer effects at high redshifts, we can simply assume that \lya\ line width is positively correlated with $N_{\mathrm{HI}}$ as $\mathrm{FWHM} \simeq 320 \left(\frac{N_{\mathrm{HI}}}{10^{20} \mathrm{cm}^{-2}}\right)^{1/3}$ \citep{Dijkstra17, Li22} for a static shell with temperature $\mathrm{T}_{\mathrm{eff}} = 10^4$ K. This translates to a column density $\mathrm{log}(N_{\mathrm{HI}}) \sim 20.18 \, \mathrm{cm}^{-2} $ for the highest FWHM LAE in our sample (i.e. MAGPI$1204117107$ at $z = 6.046$). Further, using the correlation between velocity offset and $\mathrm{log}(N_{\mathrm{HI}})$ \citep[which is approximated by a second order polynomial, see Fig. 2 of ][]{Verhamme15}, we expect this line to be at an offset of $300 \, \mathrm{km}\, \mathrm{s}^{-1}$. We follow the IGM transmission model of \cite{Witstok24} and find that this LAE can have $> 30$ \% IGM transmission if it resides inside a bubble of radius $R_{B} = 0.86$ pMpc. This estimation is in agreement with the findings of \cite{Endsley22_2}, where they discover boosted \lya\ transmission in LAEs with large velocity offsets and broad line widths. On the other hand, MAGPI$2310245276$ at $z= 6.039$ (FWHM $= 199.64 \, \mathrm{km}\, \mathrm{s}^{-1}$) can only transmit $< 10$ \% of \lya\ photons if it resides in $R_{B} = 0.39$ pMpc bubble. However, it is also possible that for a very low velocity offset ($< 200 \, \mathrm{km}\, \mathrm{s}^{-1} $), a sufficiently large ($R_{B} > 1.5$  pMpc) bubble will be able to transmit $> 60$ \% of \lya\ photons \citep[see, Fig. 3 of][]{Witstok24} by reducing the scattering effect, which can increase the observed \lya\ luminosity and may allow some photons to escape on the bluer side. A better understanding of $N_{\mathrm{HI}}$ - FWHM - velocity offset relation at $z > 6$ will require efficient radiative transfer modeling of these high-redshift LAEs.  

Our analysis suggests that galaxies with broader \lya\ emission lines are generally located within larger ionized bubbles. However, we caution that this may not always hold true. For instance, UV-bright galaxies in the reionization era may exhibit wider \lya\ line profiles if they have more \hi\ present  in their CGM, resulting in more scattering of \lya\ photons \citep[see,][]{Tang23}. Moreover, UV-bright galaxies are found to display larger \lya\ velocity offsets \citep[e.g.][]{Endsley22_2, Valentino22, Tang23, Witten23}. In such cases, most of the \lya\ photons emerge at relatively high velocities, experiencing minimal attenuation by the neutral IGM \citep[see,][]{MG20}. Visibility of \lya\ during reionization can also be enhanced by efficient ionizing photon production due to hard ionizing field \citep[e.g.][]{Mainali18, Tang23}, or if \lya\ is produced in gas clumps moving with large peculiar motions, leading to intrinsically broad \lya\ profile \citep[e.g.][]{Endsley22_2}. Therefore, wide \lya\ lines from LAEs can survive even in moderate-sized ionized bubbles \citep[e.g.][]{Endsley22_2}.  Future observations of \lya\ in both UV-bright and faint galaxies during the reionization era will provide more precise constraints on bubble sizes and the influence of the IGM on \lya\ transmission.

%\textcolor{red}{One of the main conclusions of this paper is that galaxies with wider Lyα line tend to be situated in a larger ionized bubble. However, the authors should notice that this may not always be the case. For example, Lyα line profiles tend to be wider in more UV-luminous galaxies as Lyα photons experience more neutral hydrogen in the CGM and more scatterings, and the Lyα velocity offsets are also larger. In the case, the majority of Lyα photons emerging from such systems are at relatively large velocities and the attenuation by the neutral IGM is small (e.g., Mason \& Gronke 2020; Endsley et al. 2022). Therefore, the wide Lyα emission can transmit through the neutral IGM even the ionized bubble size is small (e.g., Tang et al. 2023). I recommend the authors to moderate this conclusion, or at least discuss the scenario that wide Lyα emission can also survive in small ionized bubbles.
%}

%\begin{figure}[h!]
%    \centering
%    \includegraphics[width=1\textwidth]{Images/Rbvsskew.png}
%    \caption{Radius of ionized bubbles created by MAGPI LAEs are plotted against the asymmetry parameter of the \lya\ line.  }
%    \label{fig5}
%\end{figure}

\subsection{Closely separated LAE pair: overlapping bubbles?}\label{4.3}

We find two closely separated LAE pairs. There are two LAEs in MAGPI$2310$ field (MAGPI IDs: $2310245276$ and $2310222098$) at $z = 6.039$ and $z = 6.1485$ respectively, are located $36''$ apart on the sky, which corresponds to a physical distance of $203$ kpc. Their separation along the line of sight is $6.32$ pMpc. As a result, the estimated radii of their bubbles indicate that these bubbles do not overlap. We also find two LAEs at $z = 6.046$ in MAGPI$1529$ field (MAGPI IDs: $1529106057$ and $1529110045$), whose redshifts differ only by $0.0002$, which corresponds to a velocity shift of only $ 61 \, \mathrm{km} \, \mathrm{s}^{-1}$. They are located only $3''$ apart on the sky (see Fig.\,\ref{f5}), translating to a projected physical separation of $15.92 \,$ kpc. The estimated bubble radii for them suggest that their bubbles overlap and they likely sit inside a single large ionized region (see Fig. \ref{f5}, right panel). 

Closely separated LAEs have been previously observed in the reionization era \citep[see,][]{Castellano16, Hu17, Tilvi20, Endsley1, Endsley22, Chen24, Witstok24}. Many of these LAEs are found to reside in overdense regions populated by fainter galaxies \citep[e.g.][]{Castellano16, Castellano18, Chen24, Witstok24, Whitler24}. On the other hand, \cite{Endsley22} detect many UV-bright galaxies making an overdense region around three closely-separated Lyman-break galaxies at $z \sim 6.8$. Closely separated LAEs in overdense regions can cause bubble overlap, which results in the formation of giant ionized bubbles, allowing a boosted transmission of \lya\ through the neutral IGM \citep{Mason18a, Jung22b}. With our current data, we can not yet fully characterize the extent of the ionized regions surrounding the closely-separated LAEs in our sample. Future spectroscopic follow-up of these LAEs, as well as spectroscopy and deep photometry focusing on fainter galaxies in the surrounding area are necessary to place further constraints on the size of the ionized bubbles in those regions. Further, we note that our LAE pair at $z = 6.046$ is UV-bright ($\mathrm{M}_{\mathrm{UV}} \sim -21$), similar to that of the closely separated LAEs discussed in \cite{Endsley1}, potentially indicating that UV-bright galaxies in overdense regions enhance \lya\ transmission \citep{Endsley1, Endsley22}. Reionization simulation has also predicted that reionization is more advanced in high-density regions compared to those that are isolated, indicating that LAE clustering could lead to \lya\ emission that is less-attenuated by the IGM \citep{Qin22, Lu24}. Considering different models for the internal velocity structure of a galaxy (i.e. expanding shell or cloud), \cite{Yajima18} find that the overlapping bubbles are likely to affect the outflow velocity. Further, LAE clustering during reionization supports an accelerated reionization scenario, suggesting that reionization proceeds faster in regions around such galaxies \citep[see,][]{Endsley1}.

%\begin{figure}[h!]
%    \centering
%    \includegraphics[width=1\textwidth]{Images/Couple_final.png}
%    \caption{Synthetic MUSE \lya\ narrow band (NB) image collapsed for wavelengths with $8550$ \AA\ to $8590$ \AA, showing two LAEs with MAGPI IDs 1529106057 and 1529110045 at $z = 6.046$, separated by a physical distance of $15.92$ kpc. Contours are shown as black dashed lines at the $2$ and $4 \sigma$ significance levels. }
%    \label{f5}
%\end{figure}

\section{Summary and conclusions}\label{sec:summary}

We present spectroscopic properties of $22$ newly discovered LAEs at the edge of reionization era ($5.5 \lesssim z \lesssim 6.6$), identified in the MAGPI data. For $17$ of them covered in the HSC-Subaru Wide layer, we provide photometric magnitudes and $2\sigma$ limits (for non-detections) of i, z, y broad-band filters. The HSC-y band magnitudes indicate that our LAEs are UV-bright, with rest-frame absolute UV magnitudes $ -19.74 \leq \mathrm{M}_{\mathrm{UV}} \leq -23.27$. We summarize our main findings as follows: 

i) We observe that for $z < 6$, the FWHM distribution of LAEs remains almost uniform with no significant change as luminosity increases. However, for $z > 6$, there is a noticeable increase in line width with increasing luminosity. This broadening at high redshifts suggests that high-luminosity LAEs at $z > 6$ may be located in more highly ionized regions of the IGM, despite the fact that a more neutral IGM would typically narrow the line by scattering more \lya\ photons. 

ii) We also find that some low-equivalent width ($\mathrm{EW}_0 < 20$ \AA) LAEs exhibit higher ionizing photon production efficiency. This suggests a possible anti-correlation between the \lya\ escape fraction and $\xi_{\mathrm{ion}}$, as observed in other studies. This anti-correlation might indicate a time delay between the production and escape of ionizing photons in these galaxies, potentially related to supernova activity.

iii) Using the \lya\ spectroscopic properties and based on some assumptions, we find that LAEs in our sample can ionize bubbles of size $R_{\mathrm{B}} = 0.31 - 0.86$ pMpc. The study also finds a significant linear correlation between bubble radii and \lya\ line widths. The correlation is particularly strong in LAEs at $z > 6$, supporting the idea that the broadening of \lya\ lines at these high redshifts is due to large ionized bubbles around those galaxies. Larger bubbles allow most of the \lya\ photons to escape the galaxy with minimal scattering in the IGM. Based on simple model and assumptions, \lya\ line width can be used to constrain the IGM transmission as well as bubble sizes. A narrow line width could indicate low \hi\ column density, which will force \lya\ photons to escape close to the line centre, where line flux is significantly reduced due to IGM scattering. However, this can be uncertain and depends on the complex radiative transfer of \lya\ photons in the IGM and CGM. We also discuss that the enhanced visibility of \lya\ could also be due to several other factors such as large velocity offset of \lya\ form systemic velocity, hardness of ionizing field and due to large peculiar motions of gas clumps present in the galaxy. In such cases, wide \lya\ can transmit through the neutral IGM even when the ionized bubble size is small.  

iv) Two closely-separated LAE pairs are discovered in two MAGPI fields. One pair at $z = 6.046$ is just $3''$ apart on the sky, which corresponds to a physical separation of about $15.92$ kpc. Their bubble radii indicate that they sit inside a single large ionized region. As far as we know, this is the pair of LAEs with the smallest separation ever identified in the reionization era. Such overlap of ionized bubbles due to clustering of LAEs during reionization increases the size of the ionized regions and enhances the transmission of \lya\ through the neutral IGM. Future spectroscopic and photometric follow-up of the area surrounding these LAEs will detect fainter galaxies, allowing for improved constraints on the size of the ionized bubbles in those regions.

This study shows how wide-area \lya\ spectroscopy across a broad range of redshifts can aid in identifying and characterizing ionized bubbles formed during reionization. In the future, as new reionization-era LAEs will be discovered in the upcoming MAGPI fields, it will be possible to place more precise constraints on the evolution of line width with luminosity. Additionally, spectroscopic follow-up using infrared spectroscopy will allow for the measurement of systemic redshifts, leading to more accurate determinations of bubble sizes.

\begin{acknowledgement}

We thank the anonymous referee for their constructive feedback that helped to improve this work. We wish to thank the ESO staff, and in particular the staff at Paranal Observatory, for carrying out the MAGPI observations. MAGPI targets were selected from GAMA. GAMA is a joint European-Australasian project based around a spectroscopic campaign using the Anglo-Australian Telescope. GAMA was funded by the STFC (UK), the ARC (Australia), the AAO, and the participating institutions. GAMA photometry is based on observations made with ESO Telescopes at the La Silla Paranal Observatory under programme ID 179.A-2004, ID 177.A-3016. The MAGPI team acknowledge support from the Australian Research Council Centre of Excellence for All Sky Astrophysics in 3 Dimensions (ASTRO 3D), through project number CE170100013.

The Hyper Suprime-Cam (HSC) collaboration includes the astronomical communities of Japan and Taiwan, and Princeton University. The HSC instrumentation and software were developed by the National Astronomical Observatory of Japan (NAOJ), the Kavli Institute for the Physics and Mathematics of the Universe (Kavli IPMU), the University of Tokyo, the High Energy Accelerator Research Organization (KEK), the Academia Sinica Institute for Astronomy and Astrophysics in Taiwan (ASIAA), and Princeton University. Funding was contributed by the FIRST program from Japanese Cabinet Office, the Ministry of Education, Culture, Sports, Science and Technology (MEXT), the Japan Society for the Promotion of Science (JSPS), Japan Science and Technology Agency (JST), the Toray Science Foundation, NAOJ, Kavli IPMU, KEK, ASIAA, and Princeton University. 

This paper is based on data collected at the Subaru Telescope and retrieved from the HSC data archive system, which is operated by the Subaru Telescope and Astronomy Data Center (ADC) at NAOJ. Data analysis was in part carried out with the cooperation of Center for Computational Astrophysics (CfCA), NAOJ.

SGL acknowledge the financial support from the MICIU with funding from the European Union NextGenerationEU and Generalitat Valenciana in the call Programa de 
Planes Complementarios de I+D+i (PRTR 2022) Project (VAL-JPAS), reference ASFAE/2022/025. KEH acknowledges funding from the Australian Research Council (ARC) Discovery Project DP210101945. SMS acknowledges funding from the Australian Research Council (DE220100003). Parts of this research were conducted by the Australian Research Council Centre of Excellence for All Sky Astrophysics in 3 Dimensions (ASTRO 3D), through project number CE170100013. LMV acknowledges support by the German Academic Scholarship Foundation (Studienstiftung des deutschen Volkes) and the 
Marianne-Plehn-Program of the Elite Network of Bavaria.

\end{acknowledgement}

%\endnote in some journals will behave like \footnote; and \printendnotes will not output anything. 
\printendnotes
%\addbibresource{pasa.bib}
\bibliography{pasa}

\begin{thebibliography}{}
\expandafter\ifx\csname natexlab\endcsname\relax\def\natexlab#1{#1}\fi

\bibitem[{{Aihara} {et~al.}(2018){Aihara}, {Arimoto}, {Armstrong}, {Arnouts}, {Bahcall}, {Bickerton}, {Bosch}, {Bundy}, {Capak}, {Chan}, {Chiba}, {Coupon}, {Egami}, {Enoki}, {Finet}, {Fujimori}, {Fujimoto}, {Furusawa}, {Furusawa}, {Goto}, {Goulding}, {Greco}, {Greene}, {Gunn}, {Hamana}, {Harikane}, {Hashimoto}, {Hattori}, {Hayashi}, {Hayashi}, {He{\l}miniak}, {Higuchi}, {Hikage}, {Ho}, {Hsieh}, {Huang}, {Huang}, {Ikeda}, {Imanishi}, {Inoue}, {Iwasawa}, {Iwata}, {Jaelani}, {Jian}, {Kamata}, {Karoji}, {Kashikawa}, {Katayama}, {Kawanomoto}, {Kayo}, {Koda}, {Koike}, {Kojima}, {Komiyama}, {Konno}, {Koshida}, {Koyama}, {Kusakabe}, {Leauthaud}, {Lee}, {Lin}, {Lin}, {Lupton}, {Mandelbaum}, {Matsuoka}, {Medezinski}, {Mineo}, {Miyama}, {Miyatake}, {Miyazaki}, {Momose}, {More}, {More}, {Moritani}, {Moriya}, {Morokuma}, {Mukae}, {Murata}, {Murayama}, {Nagao}, {Nakata}, {Niida}, {Niikura}, {Nishizawa}, {Obuchi}, {Oguri}, {Oishi}, {Okabe}, {Okamoto}, {Okura}, {Ono}, {Onodera}, {Onoue}, {Osato}, {Ouchi}, {Price}, {Pyo},
  {Sako}, {Sawicki}, {Shibuya}, {Shimasaku}, {Shimono}, {Shirasaki}, {Silverman}, {Simet}, {Speagle}, {Spergel}, {Strauss}, {Sugahara}, {Sugiyama}, {Suto}, {Suyu}, {Suzuki}, {Tait}, {Takada}, {Takata}, {Tamura}, {Tanaka}, {Tanaka}, {Tanaka}, {Tanaka}, {Terai}, {Terashima}, {Toba}, {Tominaga}, {Toshikawa}, {Turner}, {Uchida}, {Uchiyama}, {Umetsu}, {Uraguchi}, {Urata}, {Usuda}, {Utsumi}, {Wang}, {Wang}, {Wong}, {Yabe}, {Yamada}, {Yamanoi}, {Yasuda}, {Yeh}, {Yonehara}, \& {Yuma}}]{Aihara18}
{Aihara}, H., {Arimoto}, N., {Armstrong}, R., {et~al.} 2018, \pasj, 70, S4

\bibitem[{{Aihara} {et~al.}(2019){Aihara}, {AlSayyad}, {Ando}, {Armstrong}, {Bosch}, {Egami}, {Furusawa}, {Furusawa}, {Goulding}, {Harikane}, {Hikage}, {Ho}, {Hsieh}, {Huang}, {Ikeda}, {Imanishi}, {Ito}, {Iwata}, {Jaelani}, {Kakuma}, {Kawana}, {Kikuta}, {Kobayashi}, {Koike}, {Komiyama}, {Li}, {Liang}, {Lin}, {Luo}, {Lupton}, {Lust}, {MacArthur}, {Matsuoka}, {Mineo}, {Miyatake}, {Miyazaki}, {More}, {Murata}, {Namiki}, {Nishizawa}, {Oguri}, {Okabe}, {Okamoto}, {Okura}, {Ono}, {Onodera}, {Onoue}, {Osato}, {Ouchi}, {Shibuya}, {Strauss}, {Sugiyama}, {Suto}, {Takada}, {Takagi}, {Takata}, {Takita}, {Tanaka}, {Terai}, {Toba}, {Uchiyama}, {Utsumi}, {Wang}, {Wang}, \& {Yamada}}]{Aihara19}
{Aihara}, H., {AlSayyad}, Y., {Ando}, M., {et~al.} 2019, \pasj, 71, 114

\bibitem[{{Bacon} {et~al.}(2010){Bacon}, {Accardo}, {Adjali}, {Anwand}, {Bauer}, {Biswas}, {Blaizot}, {Boudon}, {Brau-Nogue}, {Brinchmann}, {Caillier}, {Capoani}, {Carollo}, {Contini}, {Couderc}, {Daguis{\'e}}, {Deiries}, {Delabre}, {Dreizler}, {Dubois}, {Dupieux}, {Dupuy}, {Emsellem}, {Fechner}, {Fleischmann}, {Fran{\c{c}}ois}, {Gallou}, {Gharsa}, {Glindemann}, {Gojak}, {Guiderdoni}, {Hansali}, {Hahn}, {Jarno}, {Kelz}, {Koehler}, {Kosmalski}, {Laurent}, {Le Floch}, {Lilly}, {Lizon}, {Loupias}, {Manescau}, {Monstein}, {Nicklas}, {Olaya}, {Pares}, {Pasquini}, {P{\'e}contal-Rousset}, {Pell{\'o}}, {Petit}, {Popow}, {Reiss}, {Remillieux}, {Renault}, {Roth}, {Rupprecht}, {Serre}, {Schaye}, {Soucail}, {Steinmetz}, {Streicher}, {Stuik}, {Valentin}, {Vernet}, {Weilbacher}, {Wisotzki}, \& {Yerle}}]{Bacon10}
{Bacon}, R., {Accardo}, M., {Adjali}, L., {et~al.} 2010, in Society of Photo-Optical Instrumentation Engineers (SPIE) Conference Series, Vol. 7735, Ground-based and Airborne Instrumentation for Astronomy III, ed. I.~S. {McLean}, S.~K. {Ramsay}, \& H.~{Takami}, 773508

\bibitem[{{Bacon} {et~al.}(2023){Bacon}, {Brinchmann}, {Conseil}, {Maseda}, {Nanayakkara}, {Wendt}, {Bacher}, {Mary}, {Weilbacher}, {Krajnovi{\'c}}, {Boogaard}, {Bouch{\'e}}, {Contini}, {Epinat}, {Feltre}, {Guo}, {Herenz}, {Kollatschny}, {Kusakabe}, {Leclercq}, {Michel-Dansac}, {Pello}, {Richard}, {Roth}, {Salvignol}, {Schaye}, {Steinmetz}, {Tresse}, {Urrutia}, {Verhamme}, {Vitte}, {Wisotzki}, \& {Zoutendijk}}]{Bacon23}
{Bacon}, R., {Brinchmann}, J., {Conseil}, S., {et~al.} 2023, \aap, 670, A4

\bibitem[{{Bagley} {et~al.}(2017){Bagley}, {Scarlata}, {Henry}, {Rafelski}, {Malkan}, {Teplitz}, {Dai}, {Baronchelli}, {Colbert}, {Rutkowski}, {Mehta}, {Dressler}, {McCarthy}, {Bunker}, {Atek}, {Garel}, {Martin}, {Hathi}, \& {Siana}}]{Bagley17}
{Bagley}, M.~B., {Scarlata}, C., {Henry}, A., {et~al.} 2017, \apj, 837, 11

\bibitem[{{Becker} {et~al.}(2015){Becker}, {Bolton}, {Madau}, {Pettini}, {Ryan-Weber}, \& {Venemans}}]{Becker15}
{Becker}, G.~D., {Bolton}, J.~S., {Madau}, P., {et~al.} 2015, \mnras, 447, 3402

\bibitem[{{Begley} {et~al.}(2024){Begley}, {Cullen}, {McLure}, {Shapley}, {Dunlop}, {Carnall}, {McLeod}, {Donnan}, {Hamadouche}, \& {Stanton}}]{Begley24}
{Begley}, R., {Cullen}, F., {McLure}, R.~J., {et~al.} 2024, \mnras, 527, 4040

\bibitem[{{Bosman} {et~al.}(2022){Bosman}, {Davies}, {Becker}, {Keating}, {Davies}, {Zhu}, {Eilers}, {D'Odorico}, {Bian}, {Bischetti}, {Cristiani}, {Fan}, {Farina}, {Haehnelt}, {Hennawi}, {Kulkarni}, {Mesinger}, {Meyer}, {Onoue}, {Pallottini}, {Qin}, {Ryan-Weber}, {Schindler}, {Walter}, {Wang}, \& {Yang}}]{Bosman22}
{Bosman}, S. E.~I., {Davies}, F.~B., {Becker}, G.~D., {et~al.} 2022, \mnras, 514, 55

\bibitem[{{Bouwens} {et~al.}(2016){Bouwens}, {Smit}, {Labb{\'e}}, {Franx}, {Caruana}, {Oesch}, {Stefanon}, \& {Rasappu}}]{Bouwens16}
{Bouwens}, R.~J., {Smit}, R., {Labb{\'e}}, I., {et~al.} 2016, \apj, 831, 176

\bibitem[{{Bouwens} {et~al.}(2014){Bouwens}, {Illingworth}, {Oesch}, {Labb{\'e}}, {van Dokkum}, {Trenti}, {Franx}, {Smit}, {Gonzalez}, \& {Magee}}]{Bouwens14}
{Bouwens}, R.~J., {Illingworth}, G.~D., {Oesch}, P.~A., {et~al.} 2014, \apj, 793, 115

\bibitem[{{Bruzual} \& {Charlot}(2003)}]{Bruzual2003}
{Bruzual}, G., \& {Charlot}, S. 2003, \mnras, 344, 1000

\bibitem[{{Cain} {et~al.}(2021){Cain}, {D'Aloisio}, {Gangolli}, \& {Becker}}]{Cain21}
{Cain}, C., {D'Aloisio}, A., {Gangolli}, N., \& {Becker}, G.~D. 2021, \apjl, 917, L37

\bibitem[{{Castellano} {et~al.}(2016){Castellano}, {Dayal}, {Pentericci}, {Fontana}, {Hutter}, {Brammer}, {Merlin}, {Grazian}, {Pilo}, {Amorin}, {Cristiani}, {Dickinson}, {Ferrara}, {Gallerani}, {Giallongo}, {Giavalisco}, {Guaita}, {Koekemoer}, {Maiolino}, {Paris}, {Santini}, {Vallini}, {Vanzella}, \& {Wagg}}]{Castellano16}
{Castellano}, M., {Dayal}, P., {Pentericci}, L., {et~al.} 2016, \apjl, 818, L3

\bibitem[{{Castellano} {et~al.}(2018){Castellano}, {Pentericci}, {Vanzella}, {Marchi}, {Fontana}, {Dayal}, {Ferrara}, {Hutter}, {Carniani}, {Cristiani}, {Dickinson}, {Gallerani}, {Giallongo}, {Giavalisco}, {Grazian}, {Maiolino}, {Merlin}, {Paris}, {Pilo}, \& {Santini}}]{Castellano18}
{Castellano}, M., {Pentericci}, L., {Vanzella}, E., {et~al.} 2018, \apjl, 863, L3

\bibitem[{{Cen} \& {Haiman}(2000)}]{Cen2000}
{Cen}, R., \& {Haiman}, Z. 2000, \apjl, 542, L75

\bibitem[{{Chen} {et~al.}(2024){Chen}, {Stark}, {Mason}, {Topping}, {Whitler}, {Tang}, {Endsley}, \& {Charlot}}]{Chen24}
{Chen}, Z., {Stark}, D.~P., {Mason}, C., {et~al.} 2024, \mnras, 528, 7052

\bibitem[{{Chisholm} {et~al.}(2018){Chisholm}, {Gazagnes}, {Schaerer}, {Verhamme}, {Rigby}, {Bayliss}, {Sharon}, {Gladders}, \& {Dahle}}]{Chisholm18}
{Chisholm}, J., {Gazagnes}, S., {Schaerer}, D., {et~al.} 2018, \aap, 616, A30

\bibitem[{{Choustikov} {et~al.}(2024){Choustikov}, {Katz}, {Saxena}, {Cameron}, {Devriendt}, {Slyz}, {Rosdahl}, {Blaizot}, \& {Michel-Dansac}}]{Choustikov24}
{Choustikov}, N., {Katz}, H., {Saxena}, A., {et~al.} 2024, \mnras, 529, 3751

\bibitem[{{Claeyssens} {et~al.}(2019){Claeyssens}, {Richard}, {Blaizot}, {Garel}, {Leclercq}, {Patr{\'\i}cio}, {Verhamme}, {Wisotzki}, {Bacon}, {Carton}, {Cl{\'e}ment}, {Herenz}, {Marino}, {Muzahid}, {Saust}, \& {Schaye}}]{Claeyssens19}
{Claeyssens}, A., {Richard}, J., {Blaizot}, J., {et~al.} 2019, \mnras, 489, 5022

\bibitem[{{Dayal} {et~al.}(2020){Dayal}, {Volonteri}, {Choudhury}, {Schneider}, {Trebitsch}, {Gnedin}, {Atek}, {Hirschmann}, \& {Reines}}]{Dayal20}
{Dayal}, P., {Volonteri}, M., {Choudhury}, T.~R., {et~al.} 2020, \mnras, 495, 3065

\bibitem[{{Dijkstra}(2017)}]{Dijkstra17}
{Dijkstra}, M. 2017, arXiv e-prints, arXiv:1704.03416

\bibitem[{{Dijkstra} {et~al.}(2011){Dijkstra}, {Mesinger}, \& {Wyithe}}]{Dijkstra11}
{Dijkstra}, M., {Mesinger}, A., \& {Wyithe}, J. S.~B. 2011, \mnras, 414, 2139

\bibitem[{{Driver} {et~al.}(2011){Driver}, {Hill}, {Kelvin}, {Robotham}, {Liske}, {Norberg}, {Baldry}, {Bamford}, {Hopkins}, {Loveday}, {Peacock}, {Andrae}, {Bland-Hawthorn}, {Brough}, {Brown}, {Cameron}, {Ching}, {Colless}, {Conselice}, {Croom}, {Cross}, {de Propris}, {Dye}, {Drinkwater}, {Ellis}, {Graham}, {Grootes}, {Gunawardhana}, {Jones}, {van Kampen}, {Maraston}, {Nichol}, {Parkinson}, {Phillipps}, {Pimbblet}, {Popescu}, {Prescott}, {Roseboom}, {Sadler}, {Sansom}, {Sharp}, {Smith}, {Taylor}, {Thomas}, {Tuffs}, {Wijesinghe}, {Dunne}, {Frenk}, {Jarvis}, {Madore}, {Meyer}, {Seibert}, {Staveley-Smith}, {Sutherland}, \& {Warren}}]{Driver11}
{Driver}, S.~P., {Hill}, D.~T., {Kelvin}, L.~S., {et~al.} 2011, \mnras, 413, 971

\bibitem[{{Dunlop} {et~al.}(2013){Dunlop}, {Rogers}, {McLure}, {Ellis}, {Robertson}, {Koekemoer}, {Dayal}, {Curtis-Lake}, {Wild}, {Charlot}, {Bowler}, {Schenker}, {Ouchi}, {Ono}, {Cirasuolo}, {Furlanetto}, {Stark}, {Targett}, \& {Schneider}}]{Dunlop13}
{Dunlop}, J.~S., {Rogers}, A.~B., {McLure}, R.~J., {et~al.} 2013, \mnras, 432, 3520

\bibitem[{{Endsley} \& {Stark}(2022)}]{Endsley22}
{Endsley}, R., \& {Stark}, D.~P. 2022, \mnras, 511, 6042

\bibitem[{{Endsley} {et~al.}(2021{\natexlab{a}}){Endsley}, {Stark}, {Charlot}, {Chevallard}, {Robertson}, {Bouwens}, \& {Stefanon}}]{Endsley1}
{Endsley}, R., {Stark}, D.~P., {Charlot}, S., {et~al.} 2021{\natexlab{a}}, \mnras, 502, 6044

\bibitem[{{Endsley} {et~al.}(2021{\natexlab{b}}){Endsley}, {Stark}, {Chevallard}, \& {Charlot}}]{Endsley2}
{Endsley}, R., {Stark}, D.~P., {Chevallard}, J., \& {Charlot}, S. 2021{\natexlab{b}}, \mnras, 500, 5229

\bibitem[{{Endsley} {et~al.}(2022){Endsley}, {Stark}, {Bouwens}, {Schouws}, {Smit}, {Stefanon}, {Inami}, {Bowler}, {Oesch}, {Gonzalez}, {Aravena}, {da Cunha}, {Dayal}, {Ferrara}, {Graziani}, {Nanayakkara}, {Pallottini}, {Schneider}, {Sommovigo}, {Topping}, {van der Werf}, \& {Hutter}}]{Endsley22_2}
{Endsley}, R., {Stark}, D.~P., {Bouwens}, R.~J., {et~al.} 2022, \mnras, 517, 5642

\bibitem[{{Fan} {et~al.}(2006){Fan}, {Strauss}, {Becker}, {White}, {Gunn}, {Knapp}, {Richards}, {Schneider}, {Brinkmann}, \& {Fukugita}}]{Fan06}
{Fan}, X., {Strauss}, M.~A., {Becker}, R.~H., {et~al.} 2006, \aj, 132, 117

\bibitem[{{Finkelstein} {et~al.}(2019){Finkelstein}, {D'Aloisio}, {Paardekooper}, {Ryan}, {Behroozi}, {Finlator}, {Livermore}, {Upton Sanderbeck}, {Dalla Vecchia}, \& {Khochfar}}]{Finkelstein19}
{Finkelstein}, S.~L., {D'Aloisio}, A., {Paardekooper}, J.-P., {et~al.} 2019, \apj, 879, 36

\bibitem[{{Flury} {et~al.}(2022){Flury}, {Jaskot}, {Ferguson}, {Worseck}, {Makan}, {Chisholm}, {Saldana-Lopez}, {Schaerer}, {McCandliss}, {Xu}, {Wang}, {Oey}, {Ford}, {Heckman}, {Ji}, {Giavalisco}, {Amor{\'\i}n}, {Atek}, {Blaizot}, {Borthakur}, {Carr}, {Castellano}, {De Barros}, {Dickinson}, {Finkelstein}, {Fleming}, {Fontanot}, {Garel}, {Grazian}, {Hayes}, {Henry}, {Mauerhofer}, {Micheva}, {Ostlin}, {Papovich}, {Pentericci}, {Ravindranath}, {Rosdahl}, {Rutkowski}, {Santini}, {Scarlata}, {Teplitz}, {Thuan}, {Trebitsch}, {Vanzella}, \& {Verhamme}}]{Flury22}
{Flury}, S.~R., {Jaskot}, A.~E., {Ferguson}, H.~C., {et~al.} 2022, \apj, 930, 126

\bibitem[{{Foster} {et~al.}(2021){Foster}, {Mendel}, {Lagos}, {Wisnioski}, {Yuan}, {D'Eugenio}, {Barone}, {Harborne}, {Vaughan}, {Schulze}, {Remus}, {Gupta}, {Collacchioni}, {Khim}, {Taylor}, {Bassett}, {Croom}, {McDermid}, {Poci}, {Battisti}, {Bland-Hawthorn}, {Bellstedt}, {Colless}, {Davies}, {Derkenne}, {Driver}, {Ferr{\'e}-Mateu}, {Fisher}, {Gjergo}, {Johnston}, {Khalid}, {Kobayashi}, {Oh}, {Peng}, {Robotham}, {Sharda}, {Sweet}, {Taylor}, {Tran}, {Trayford}, {van de Sande}, {Yi}, \& {Zanisi}}]{Foster21}
{Foster}, C., {Mendel}, J.~T., {Lagos}, C.~D.~P., {et~al.} 2021, \pasa, 38, e031

\bibitem[{{Furtak} {et~al.}(2022){Furtak}, {Plat}, {Zitrin}, {Topping}, {Stark}, {Strait}, {Charlot}, {Coe}, {Andrade-Santos}, {Brada{\v{c}}}, {Bradley}, {Lemaux}, \& {Sharon}}]{Furtak22}
{Furtak}, L.~J., {Plat}, A., {Zitrin}, A., {et~al.} 2022, \mnras, 516, 1373

\bibitem[{{Gazagnes} {et~al.}(2020){Gazagnes}, {Chisholm}, {Schaerer}, {Verhamme}, \& {Izotov}}]{Gazagnes20}
{Gazagnes}, S., {Chisholm}, J., {Schaerer}, D., {Verhamme}, A., \& {Izotov}, Y. 2020, \aap, 639, A85

\bibitem[{{Graham} \& {Driver}(2005)}]{Graham05}
{Graham}, A.~W., \& {Driver}, S.~P. 2005, \pasa, 22, 118

\bibitem[{{Guaita} {et~al.}(2015){Guaita}, {Melinder}, {Hayes}, {{\"O}stlin}, {Gonzalez}, {Micheva}, {Adamo}, {Mas-Hesse}, {Sandberg}, {Ot{\'\i}-Floranes}, {Schaerer}, {Verhamme}, {Freeland}, {Orlitov{\'a}}, {Laursen}, {Cannon}, {Duval}, {Rivera-Thorsen}, {Herenz}, {Kunth}, {Atek}, {Puschnig}, {Gruyters}, \& {Pardy}}]{Guaita15}
{Guaita}, L., {Melinder}, J., {Hayes}, M., {et~al.} 2015, \aap, 576, A51

\bibitem[{{Gupta} {et~al.}(2024){Gupta}, {Trott}, {Jaiswar}, {Ryan-Weber}, {Bunker}, {Acharyya}, {Cameron}, {Forrest}, {Kacprzak}, {Nanayakkara}, {Tran}, \& {Chokshi}}]{Gupta24}
{Gupta}, A., {Trott}, C.~M., {Jaiswar}, R., {et~al.} 2024, \apj, 973, 169

\bibitem[{{Harikane} {et~al.}(2018){Harikane}, {Ouchi}, {Shibuya}, {Kojima}, {Zhang}, {Itoh}, {Ono}, {Higuchi}, {Inoue}, {Chevallard}, {Capak}, {Nagao}, {Onodera}, {Faisst}, {Martin}, {Rauch}, {Bruzual}, {Charlot}, {Davidzon}, {Fujimoto}, {Hilmi}, {Ilbert}, {Lee}, {Matsuoka}, {Silverman}, \& {Toft}}]{Harikane18}
{Harikane}, Y., {Ouchi}, M., {Shibuya}, T., {et~al.} 2018, \apj, 859, 84

\bibitem[{{Hashimoto} {et~al.}(2017){Hashimoto}, {Garel}, {Guiderdoni}, {Drake}, {Bacon}, {Blaizot}, {Richard}, {Leclercq}, {Inami}, {Verhamme}, {Bouwens}, {Brinchmann}, {Cantalupo}, {Carollo}, {Caruana}, {Herenz}, {Kerutt}, {Marino}, {Mitchell}, \& {Schaye}}]{Hashimoto17}
{Hashimoto}, T., {Garel}, T., {Guiderdoni}, B., {et~al.} 2017, \aap, 608, A10

\bibitem[{{Hayes} {et~al.}(2021){Hayes}, {Runnholm}, {Gronke}, \& {Scarlata}}]{Hayes21}
{Hayes}, M.~J., {Runnholm}, A., {Gronke}, M., \& {Scarlata}, C. 2021, \apj, 908, 36

\bibitem[{{Hayes} \& {Scarlata}(2023)}]{Hayes23}
{Hayes}, M.~J., \& {Scarlata}, C. 2023, \apjl, 954, L14

\bibitem[{{Henry} {et~al.}(2015){Henry}, {Scarlata}, {Martin}, \& {Erb}}]{Henry15}
{Henry}, A., {Scarlata}, C., {Martin}, C.~L., \& {Erb}, D. 2015, \apj, 809, 19

\bibitem[{{Herenz} \& {Wisotzki}(2017)}]{HW17}
{Herenz}, E.~C., \& {Wisotzki}, L. 2017, \aap, 602, A111

\bibitem[{{Herenz} {et~al.}(2017){Herenz}, {Urrutia}, {Wisotzki}, {Kerutt}, {Saust}, {Werhahn}, {Schmidt}, {Caruana}, {Diener}, {Bacon}, {Brinchmann}, {Schaye}, {Maseda}, \& {Weilbacher}}]{Herenz17}
{Herenz}, E.~C., {Urrutia}, T., {Wisotzki}, L., {et~al.} 2017, \aap, 606, A12

\bibitem[{{Hinton} {et~al.}(2016){Hinton}, {Davis}, {Lidman}, {Glazebrook}, \& {Lewis}}]{Hinton16}
{Hinton}, S.~R., {Davis}, T.~M., {Lidman}, C., {Glazebrook}, K., \& {Lewis}, G.~F. 2016, Astronomy and Computing, 15, 61

\bibitem[{{Hoag} {et~al.}(2019){Hoag}, {Brada{\v{c}}}, {Huang}, {Mason}, {Treu}, {Schmidt}, {Trenti}, {Strait}, {Lemaux}, {Finney}, \& {Paddock}}]{Hoag19}
{Hoag}, A., {Brada{\v{c}}}, M., {Huang}, K., {et~al.} 2019, \apj, 878, 12

\bibitem[{{Hu} {et~al.}(2010){Hu}, {Cowie}, {Barger}, {Capak}, {Kakazu}, \& {Trouille}}]{Hu10}
{Hu}, E.~M., {Cowie}, L.~L., {Barger}, A.~J., {et~al.} 2010, \apj, 725, 394

\bibitem[{{Hu} {et~al.}(2016){Hu}, {Cowie}, {Songaila}, {Barger}, {Rosenwasser}, \& {Wold}}]{Hu16}
{Hu}, E.~M., {Cowie}, L.~L., {Songaila}, A., {et~al.} 2016, \apjl, 825, L7

\bibitem[{{Hu} {et~al.}(2017){Hu}, {Wang}, {Zheng}, {Malhotra}, {Infante}, {Rhoads}, {Gonzalez}, {Walker}, {Jiang}, {Jiang}, {Hibon}, {Barrientos}, {Finkelstein}, {Galaz}, {Kang}, {Kong}, {Tilvi}, {Yang}, \& {Zheng}}]{Hu17}
{Hu}, W., {Wang}, J., {Zheng}, Z.-Y., {et~al.} 2017, \apjl, 845, L16

\bibitem[{{Inoue} {et~al.}(2014){Inoue}, {Shimizu}, {Iwata}, \& {Tanaka}}]{Inoue14}
{Inoue}, A.~K., {Shimizu}, I., {Iwata}, I., \& {Tanaka}, M. 2014, \mnras, 442, 1805

\bibitem[{{Izotov} {et~al.}(2020){Izotov}, {Schaerer}, {Worseck}, {Verhamme}, {Guseva}, {Thuan}, {Orlitov{\'a}}, \& {Fricke}}]{Izotov20}
{Izotov}, Y.~I., {Schaerer}, D., {Worseck}, G., {et~al.} 2020, \mnras, 491, 468

\bibitem[{{Izotov} {et~al.}(2018){Izotov}, {Worseck}, {Schaerer}, {Guseva}, {Thuan}, {Fricke}, \& {Orlitov{\'a}}}]{Izotov18}
{Izotov}, Y.~I., {Worseck}, G., {Schaerer}, D., {et~al.} 2018, \mnras, 478, 4851

\bibitem[{{Jones} {et~al.}(2024){Jones}, {Bunker}, {Saxena}, {Arribas}, {Bhatawdekar}, {Boyett}, {Carniani}, {Charlot}, {Curtis-Lake}, {Hainline}, {Johnson}, {Kumari}, {Maseda}, {Rix}, {Robertson}, {Tacchella}, {{\"U}bler}, {Williams}, {Willott}, {Witstok}, \& {Zhu}}]{Jones24}
{Jones}, G.~C., {Bunker}, A.~J., {Saxena}, A., {et~al.} 2024, arXiv e-prints, arXiv:2409.06405

\bibitem[{{Jung} {et~al.}(2022{\natexlab{a}}){Jung}, {Papovich}, {Finkelstein}, {Simons}, {Estrada-Carpenter}, {Backhaus}, {Cleri}, {Finlator}, {Giavalisco}, {Ji}, {Matharu}, {Momcheva}, {Straughn}, \& {Trump}}]{Jung22b}
{Jung}, I., {Papovich}, C., {Finkelstein}, S.~L., {et~al.} 2022{\natexlab{a}}, \apj, 933, 87

\bibitem[{{Jung} {et~al.}(2022{\natexlab{b}}){Jung}, {Finkelstein}, {Larson}, {Hutchison}, {Straughn}, {Bagley}, {Castellano}, {Cleri}, {Cooper}, {Dickinson}, {Ferguson}, {Holwerda}, {Kartaltepe}, {Kim}, {Koekemoer}, {Papovich}, {Park}, {Pentericci}, {Perez-Gonzalez}, {Song}, {Tacchella}, {Weiner}, {Willmer}, \& {Zavala}}]{Jung22a}
{Jung}, I., {Finkelstein}, S.~L., {Larson}, R.~L., {et~al.} 2022{\natexlab{b}}, arXiv e-prints, arXiv:2212.09850

\bibitem[{{Jung} {et~al.}(2024){Jung}, {Finkelstein}, {Arrabal Haro}, {Dickinson}, {Ferguson}, {Hutchison}, {Kartaltepe}, {Larson}, {Simons}, {Papovich}, {Park}, {Pentericci}, {Trump}, {Amor{\'\i}n}, {Backhaus}, {Bagley}, {Casey}, {Cheng}, {Cleri}, {Cooper}, {Cooper}, {Gardner}, {Gawiser}, {Grazian}, {Hathi}, {Hirschmann}, {Koekemoer}, {Lucas}, {Mobasher}, {Pirzkal}, {Ravindranath}, {Straughn}, {Yung}, \& {de la Vega}}]{Jung24}
{Jung}, I., {Finkelstein}, S.~L., {Arrabal Haro}, P., {et~al.} 2024, \apj, 967, 73

\bibitem[{{Kakiichi} {et~al.}(2016){Kakiichi}, {Dijkstra}, {Ciardi}, \& {Graziani}}]{Kakiichi16}
{Kakiichi}, K., {Dijkstra}, M., {Ciardi}, B., \& {Graziani}, L. 2016, \mnras, 463, 4019

\bibitem[{{Kakiichi} \& {Gronke}(2021)}]{Kakiichi21}
{Kakiichi}, K., \& {Gronke}, M. 2021, \apj, 908, 30

\bibitem[{{Katz} {et~al.}(2020){Katz}, {{\v{D}}urov{\v{c}}{\'\i}kov{\'a}}, {Kimm}, {Rosdahl}, {Blaizot}, {Haehnelt}, {Devriendt}, {Slyz}, {Ellis}, \& {Laporte}}]{Katz20}
{Katz}, H., {{\v{D}}urov{\v{c}}{\'\i}kov{\'a}}, D., {Kimm}, T., {et~al.} 2020, \mnras, 498, 164

\bibitem[{{Katz} {et~al.}(2022){Katz}, {Garel}, {Rosdahl}, {Mauerhofer}, {Kimm}, {Blaizot}, {Michel-Dansac}, {Devriendt}, {Slyz}, \& {Haehnelt}}]{Katz22}
{Katz}, H., {Garel}, T., {Rosdahl}, J., {et~al.} 2022, \mnras, 515, 4265

\bibitem[{{Kennicutt}(1998)}]{K98}
{Kennicutt}, Robert~C., J. 1998, \araa, 36, 189

\bibitem[{{Kerutt} {et~al.}(2022){Kerutt}, {Wisotzki}, {Verhamme}, {Schmidt}, {Leclercq}, {Herenz}, {Urrutia}, {Garel}, {Hashimoto}, {Maseda}, {Matthee}, {Kusakabe}, {Schaye}, {Richard}, {Guiderdoni}, {Mauerhofer}, {Nanayakkara}, \& {Vitte}}]{Kerutt22}
{Kerutt}, J., {Wisotzki}, L., {Verhamme}, A., {et~al.} 2022, \aap, 659, A183

\bibitem[{{Kerutt} {et~al.}(2024){Kerutt}, {Oesch}, {Wisotzki}, {Verhamme}, {Atek}, {Herenz}, {Illingworth}, {Kusakabe}, {Matthee}, {Mauerhofer}, {Montes}, {Naidu}, {Nelson}, {Reddy}, {Schaye}, {Simmonds}, {Urrutia}, \& {Vitte}}]{Kerutt24}
{Kerutt}, J., {Oesch}, P.~A., {Wisotzki}, L., {et~al.} 2024, \aap, 684, A42

\bibitem[{{Konno} {et~al.}(2014){Konno}, {Ouchi}, {Ono}, {Shimasaku}, {Shibuya}, {Furusawa}, {Nakajima}, {Naito}, {Momose}, {Yuma}, \& {Iye}}]{Konno14}
{Konno}, A., {Ouchi}, M., {Ono}, Y., {et~al.} 2014, \apj, 797, 16

\bibitem[{{Konno} {et~al.}(2018){Konno}, {Ouchi}, {Shibuya}, {Ono}, {Shimasaku}, {Taniguchi}, {Nagao}, {Kobayashi}, {Kajisawa}, {Kashikawa}, {Inoue}, {Oguri}, {Furusawa}, {Goto}, {Harikane}, {Higuchi}, {Komiyama}, {Kusakabe}, {Miyazaki}, {Nakajima}, \& {Wang}}]{Konno18}
{Konno}, A., {Ouchi}, M., {Shibuya}, T., {et~al.} 2018, \pasj, 70, S16

\bibitem[{{Kulkarni} {et~al.}(2019){Kulkarni}, {Keating}, {Haehnelt}, {Bosman}, {Puchwein}, {Chardin}, \& {Aubert}}]{Kulkarni19}
{Kulkarni}, G., {Keating}, L.~C., {Haehnelt}, M.~G., {et~al.} 2019, \mnras, 485, L24

\bibitem[{{Laursen} {et~al.}(2019){Laursen}, {Sommer-Larsen}, {Milvang-Jensen}, {Fynbo}, \& {Razoumov}}]{Laursen19}
{Laursen}, P., {Sommer-Larsen}, J., {Milvang-Jensen}, B., {Fynbo}, J. P.~U., \& {Razoumov}, A.~O. 2019, \aap, 627, A84

\bibitem[{{Li} \& {Gronke}(2022)}]{Li22}
{Li}, Z., \& {Gronke}, M. 2022, \mnras, 513, 5034

\bibitem[{{Lin} {et~al.}(2024){Lin}, {Cai}, {Wu}, {Li}, {Sun}, {Fan}, {Chen}, {Li}, {Bian}, {Ning}, {Jiang}, {Bruzual}, {Charlot}, \& {Chevallard}}]{Lin24}
{Lin}, X., {Cai}, Z., {Wu}, Y., {et~al.} 2024, \apjs, 272, 33

\bibitem[{{Lu} {et~al.}(2024){Lu}, {Mason}, {Hutter}, {Mesinger}, {Qin}, {Stark}, \& {Endsley}}]{Lu24}
{Lu}, T.-Y., {Mason}, C.~A., {Hutter}, A., {et~al.} 2024, \mnras, 528, 4872

\bibitem[{{Mainali} {et~al.}(2018){Mainali}, {Zitrin}, {Stark}, {Ellis}, {Richard}, {Tang}, {Laporte}, {Oesch}, \& {McGreer}}]{Mainali18}
{Mainali}, R., {Zitrin}, A., {Stark}, D.~P., {et~al.} 2018, \mnras, 479, 1180

\bibitem[{{Maji} {et~al.}(2022){Maji}, {Verhamme}, {Rosdahl}, {Garel}, {Blaizot}, {Mauerhofer}, {Pittavino}, {Victoria Feser}, {Chuniaud}, {Kimm}, {Katz}, \& {Haehnelt}}]{Maji22}
{Maji}, M., {Verhamme}, A., {Rosdahl}, J., {et~al.} 2022, \aap, 663, A66

\bibitem[{{Malhotra} \& {Rhoads}(2006)}]{MR06}
{Malhotra}, S., \& {Rhoads}, J.~E. 2006, \apjl, 647, L95

\bibitem[{{Mascia} {et~al.}(2023){Mascia}, {Pentericci}, {Calabr{\`o}}, {Treu}, {Santini}, {Yang}, {Napolitano}, {Roberts-Borsani}, {Bergamini}, {Grillo}, {Rosati}, {Vulcani}, {Castellano}, {Boyett}, {Fontana}, {Glazebrook}, {Henry}, {Mason}, {Merlin}, {Morishita}, {Nanayakkara}, {Paris}, {Roy}, {Williams}, {Wang}, {Brammer}, {Brada{\v{c}}}, {Chen}, {Kelly}, {Koekemoer}, {Trenti}, \& {Windhorst}}]{Mascia23}
{Mascia}, S., {Pentericci}, L., {Calabr{\`o}}, A., {et~al.} 2023, \aap, 672, A155

\bibitem[{{Maseda} {et~al.}(2020){Maseda}, {Bacon}, {Lam}, {Matthee}, {Brinchmann}, {Schaye}, {Labbe}, {Schmidt}, {Boogaard}, {Bouwens}, {Cantalupo}, {Franx}, {Hashimoto}, {Inami}, {Kusakabe}, {Mahler}, {Nanayakkara}, {Richard}, \& {Wisotzki}}]{Maseda20}
{Maseda}, M.~V., {Bacon}, R., {Lam}, D., {et~al.} 2020, \mnras, 493, 5120

\bibitem[{{Mason} \& {Gronke}(2020)}]{MG20}
{Mason}, C.~A., \& {Gronke}, M. 2020, \mnras, 499, 1395

\bibitem[{{Mason} {et~al.}(2018){Mason}, {Treu}, {de Barros}, {Dijkstra}, {Fontana}, {Mesinger}, {Pentericci}, {Trenti}, \& {Vanzella}}]{Mason18a}
{Mason}, C.~A., {Treu}, T., {de Barros}, S., {et~al.} 2018, \apjl, 857, L11

\bibitem[{{Matthee} {et~al.}(2017{\natexlab{a}}){Matthee}, {Sobral}, {Best}, {Khostovan}, {Oteo}, {Bouwens}, \& {R{\"o}ttgering}}]{Matthee17a}
{Matthee}, J., {Sobral}, D., {Best}, P., {et~al.} 2017{\natexlab{a}}, \mnras, 465, 3637

\bibitem[{{Matthee} {et~al.}(2017{\natexlab{b}}){Matthee}, {Sobral}, {Darvish}, {Santos}, {Mobasher}, {Paulino-Afonso}, {R{\"o}ttgering}, \& {Alegre}}]{Matthee17}
{Matthee}, J., {Sobral}, D., {Darvish}, B., {et~al.} 2017{\natexlab{b}}, \mnras, 472, 772

\bibitem[{{Matthee} {et~al.}(2015){Matthee}, {Sobral}, {Santos}, {R{\"o}ttgering}, {Darvish}, \& {Mobasher}}]{Matthee15}
{Matthee}, J., {Sobral}, D., {Santos}, S., {et~al.} 2015, \mnras, 451, 400

\bibitem[{{Matthee} {et~al.}(2022){Matthee}, {Naidu}, {Pezzulli}, {Gronke}, {Sobral}, {Oesch}, {Hayes}, {Erb}, {Schaerer}, {Amor{\'\i}n}, {Tacchella}, {Paulino-Afonso}, {Llerena}, {Calhau}, \& {R{\"o}ttgering}}]{Matthee22}
{Matthee}, J., {Naidu}, R.~P., {Pezzulli}, G., {et~al.} 2022, \mnras, 512, 5960

\bibitem[{{Meyer} {et~al.}(2021){Meyer}, {Laporte}, {Ellis}, {Verhamme}, \& {Garel}}]{Meyer21}
{Meyer}, R.~A., {Laporte}, N., {Ellis}, R.~S., {Verhamme}, A., \& {Garel}, T. 2021, \mnras, 500, 558

\bibitem[{{Miyazaki} {et~al.}(2018){Miyazaki}, {Komiyama}, {Kawanomoto}, {Doi}, {Furusawa}, {Hamana}, {Hayashi}, {Ikeda}, {Kamata}, {Karoji}, {Koike}, {Kurakami}, {Miyama}, {Morokuma}, {Nakata}, {Namikawa}, {Nakaya}, {Nariai}, {Obuchi}, {Oishi}, {Okada}, {Okura}, {Tait}, {Takata}, {Tanaka}, {Tanaka}, {Terai}, {Tomono}, {Uraguchi}, {Usuda}, {Utsumi}, {Yamada}, {Yamanoi}, {Aihara}, {Fujimori}, {Mineo}, {Miyatake}, {Oguri}, {Uchida}, {Tanaka}, {Yasuda}, {Takada}, {Murayama}, {Nishizawa}, {Sugiyama}, {Chiba}, {Futamase}, {Wang}, {Chen}, {Ho}, {Liaw}, {Chiu}, {Ho}, {Lai}, {Lee}, {Jeng}, {Iwamura}, {Armstrong}, {Bickerton}, {Bosch}, {Gunn}, {Lupton}, {Loomis}, {Price}, {Smith}, {Strauss}, {Turner}, {Suzuki}, {Miyazaki}, {Muramatsu}, {Yamamoto}, {Endo}, {Ezaki}, {Ito}, {Kawaguchi}, {Sofuku}, {Taniike}, {Akutsu}, {Dojo}, {Kasumi}, {Matsuda}, {Imoto}, {Miwa}, {Suzuki}, {Takeshi}, \& {Yokota}}]{Miyazaki18}
{Miyazaki}, S., {Komiyama}, Y., {Kawanomoto}, S., {et~al.} 2018, \pasj, 70, S1

\bibitem[{{Mukherjee} {et~al.}(2023){Mukherjee}, {Zafar}, {Nanayakkara}, {Wisnioski}, {Battisti}, {Gupta}, {Lagos}, {Harborne}, {Foster}, {Mendel}, {Croom}, {Mailvaganam}, \& {Prathap}}]{Mukherjee23}
{Mukherjee}, T., {Zafar}, T., {Nanayakkara}, T., {et~al.} 2023, \aap, 680, L5

\bibitem[{{Naidu} {et~al.}(2020){Naidu}, {Tacchella}, {Mason}, {Bose}, {Oesch}, \& {Conroy}}]{Naidu20}
{Naidu}, R.~P., {Tacchella}, S., {Mason}, C.~A., {et~al.} 2020, \apj, 892, 109

\bibitem[{{Naidu} {et~al.}(2022){Naidu}, {Matthee}, {Oesch}, {Conroy}, {Sobral}, {Pezzulli}, {Hayes}, {Erb}, {Amor{\'\i}n}, {Gronke}, {Schaerer}, {Tacchella}, {Kerutt}, {Paulino-Afonso}, {Calhau}, {Llerena}, \& {R{\"o}ttgering}}]{Naidu22}
{Naidu}, R.~P., {Matthee}, J., {Oesch}, P.~A., {et~al.} 2022, \mnras, 510, 4582

\bibitem[{{Nakajima} {et~al.}(2020){Nakajima}, {Ellis}, {Robertson}, {Tang}, \& {Stark}}]{Nakajima20}
{Nakajima}, K., {Ellis}, R.~S., {Robertson}, B.~E., {Tang}, M., \& {Stark}, D.~P. 2020, \apj, 889, 161

\bibitem[{{Nakane} {et~al.}(2024){Nakane}, {Ouchi}, {Nakajima}, {Harikane}, {Ono}, {Umeda}, {Isobe}, {Zhang}, \& {Xu}}]{Nakane24}
{Nakane}, M., {Ouchi}, M., {Nakajima}, K., {et~al.} 2024, \apj, 967, 28

\bibitem[{{Napolitano} {et~al.}(2024){Napolitano}, {Pentericci}, {Santini}, {Calabr{\`o}}, {Mascia}, {Llerena}, {Castellano}, {Dickinson}, {Finkelstein}, {Amor{\'\i}n}, {Arrabal Haro}, {Bagley}, {Bhatawdekar}, {Cleri}, {Davis}, {Gardner}, {Gawiser}, {Giavalisco}, {Hathi}, {Holwerda}, {Hu}, {Jung}, {Kartaltepe}, {Koekemoer}, {Larson}, {Merlin}, {Mobasher}, {Papovich}, {Park}, {Pirzkal}, {Trump}, {Wilkins}, \& {Yung}}]{Napolitano24}
{Napolitano}, L., {Pentericci}, L., {Santini}, P., {et~al.} 2024, \aap, 688, A106

\bibitem[{{Neufeld}(1990)}]{Neufeld90}
{Neufeld}, D.~A. 1990, \apj, 350, 216

\bibitem[{{Ning} {et~al.}(2023){Ning}, {Cai}, {Jiang}, {Lin}, {Fu}, \& {Spinoso}}]{Ning23}
{Ning}, Y., {Cai}, Z., {Jiang}, L., {et~al.} 2023, \apjl, 944, L1

\bibitem[{{Pentericci} {et~al.}(2011){Pentericci}, {Fontana}, {Vanzella}, {Castellano}, {Grazian}, {Dijkstra}, {Boutsia}, {Cristiani}, {Dickinson}, {Giallongo}, {Giavalisco}, {Maiolino}, {Moorwood}, {Paris}, \& {Santini}}]{Pentericci11}
{Pentericci}, L., {Fontana}, A., {Vanzella}, E., {et~al.} 2011, \apj, 743, 132

\bibitem[{{Pentericci} {et~al.}(2014){Pentericci}, {Vanzella}, {Fontana}, {Castellano}, {Treu}, {Mesinger}, {Dijkstra}, {Grazian}, {Brada{\v{c}}}, {Conselice}, {Cristiani}, {Dunlop}, {Galametz}, {Giavalisco}, {Giallongo}, {Koekemoer}, {McLure}, {Maiolino}, {Paris}, \& {Santini}}]{Pentericci14}
{Pentericci}, L., {Vanzella}, E., {Fontana}, A., {et~al.} 2014, \apj, 793, 113

\bibitem[{{Prieto-Lyon} {et~al.}(2023){Prieto-Lyon}, {Strait}, {Mason}, {Brammer}, {Caminha}, {Mercurio}, {Acebron}, {Bergamini}, {Grillo}, {Rosati}, {Vanzella}, {Castellano}, {Merlin}, {Paris}, {Boyett}, {Calabr{\`o}}, {Morishita}, {Mascia}, {Pentericci}, {Roberts-Borsani}, {Roy}, {Treu}, \& {Vulcani}}]{Lyon23}
{Prieto-Lyon}, G., {Strait}, V., {Mason}, C.~A., {et~al.} 2023, \aap, 672, A186

\bibitem[{{Pucha} {et~al.}(2022){Pucha}, {Reddy}, {Dey}, {Juneau}, {Lee}, {Prescott}, {Shivaei}, \& {Hong}}]{Pucha22}
{Pucha}, R., {Reddy}, N.~A., {Dey}, A., {et~al.} 2022, \aj, 164, 159

\bibitem[{{Qin} {et~al.}(2022){Qin}, {Wyithe}, {Oesch}, {Illingworth}, {Leonova}, {Mutch}, \& {Naidu}}]{Qin22}
{Qin}, Y., {Wyithe}, J. S.~B., {Oesch}, P.~A., {et~al.} 2022, \mnras, 510, 3858

\bibitem[{{Robertson} {et~al.}(2010){Robertson}, {Ellis}, {Dunlop}, {McLure}, \& {Stark}}]{Robert10}
{Robertson}, B.~E., {Ellis}, R.~S., {Dunlop}, J.~S., {McLure}, R.~J., \& {Stark}, D.~P. 2010, \nat, 468, 49

\bibitem[{{Robertson} {et~al.}(2015){Robertson}, {Ellis}, {Furlanetto}, \& {Dunlop}}]{Robertson15}
{Robertson}, B.~E., {Ellis}, R.~S., {Furlanetto}, S.~R., \& {Dunlop}, J.~S. 2015, \apjl, 802, L19

\bibitem[{{Roy} {et~al.}(2023){Roy}, {Henry}, {Treu}, {Jones}, {Prieto-Lyon}, {Mason}, {Heckman}, {Nanayakkara}, {Pentericci}, {Mascia}, {Brada{\v{c}}}, {Vanzella}, {Scarlata}, {Boyett}, {Trenti}, \& {Wang}}]{Roy23}
{Roy}, N., {Henry}, A., {Treu}, T., {et~al.} 2023, \apjl, 952, L14

\bibitem[{{Santos} {et~al.}(2016){Santos}, {Sobral}, \& {Matthee}}]{Santos16}
{Santos}, S., {Sobral}, D., \& {Matthee}, J. 2016, \mnras, 463, 1678

\bibitem[{{Saxena} {et~al.}(2024){Saxena}, {Bunker}, {Jones}, {Stark}, {Cameron}, {Witstok}, {Arribas}, {Baker}, {Baum}, {Bhatawdekar}, {Bowler}, {Boyett}, {Carniani}, {Charlot}, {Chevallard}, {Curti}, {Curtis-Lake}, {Eisenstein}, {Endsley}, {Hainline}, {Helton}, {Johnson}, {Kumari}, {Looser}, {Maiolino}, {Rieke}, {Rix}, {Robertson}, {Sandles}, {Simmonds}, {Smit}, {Tacchella}, {Williams}, {Willmer}, \& {Willott}}]{Saxena24}
{Saxena}, A., {Bunker}, A.~J., {Jones}, G.~C., {et~al.} 2024, \aap, 684, A84

\bibitem[{{Schaerer}(2003)}]{Schaerer2003}
{Schaerer}, D. 2003, \aap, 397, 527

\bibitem[{{Schaerer} {et~al.}(2022){Schaerer}, {Izotov}, {Worseck}, {Berg}, {Chisholm}, {Jaskot}, {Nakajima}, {Ravindranath}, {Thuan}, \& {Verhamme}}]{Schaerer22}
{Schaerer}, D., {Izotov}, Y.~I., {Worseck}, G., {et~al.} 2022, \aap, 658, L11

\bibitem[{{Shibuya} {et~al.}(2014){Shibuya}, {Ouchi}, {Nakajima}, {Hashimoto}, {Ono}, {Rauch}, {Gauthier}, {Shimasaku}, {Goto}, {Mori}, \& {Umemura.}}]{Shibuya14}
{Shibuya}, T., {Ouchi}, M., {Nakajima}, K., {et~al.} 2014, \apj, 788, 74

\bibitem[{{Shibuya} {et~al.}(2018){Shibuya}, {Ouchi}, {Harikane}, {Rauch}, {Ono}, {Mukae}, {Higuchi}, {Kojima}, {Yuma}, {Lee}, {Furusawa}, {Konno}, {Martin}, {Shimasaku}, {Taniguchi}, {Kobayashi}, {Kajisawa}, {Nagao}, {Goto}, {Kashikawa}, {Komiyama}, {Kusakabe}, {Momose}, {Nakajima}, {Tanaka}, \& {Wang}}]{Shibuya18}
{Shibuya}, T., {Ouchi}, M., {Harikane}, Y., {et~al.} 2018, \pasj, 70, S15

\bibitem[{{Shivaei} {et~al.}(2018){Shivaei}, {Reddy}, {Siana}, {Shapley}, {Kriek}, {Mobasher}, {Freeman}, {Sanders}, {Coil}, {Price}, {Fetherolf}, {Azadi}, {Leung}, \& {Zick}}]{Shivaei18}
{Shivaei}, I., {Reddy}, N.~A., {Siana}, B., {et~al.} 2018, \apj, 855, 42

\bibitem[{{Simmonds} {et~al.}(2023){Simmonds}, {Tacchella}, {Maseda}, {Williams}, {Baker}, {Witten}, {Johnson}, {Robertson}, {Saxena}, {Sun}, {Witstok}, {Bhatawdekar}, {Boyett}, {Bunker}, {Charlot}, {Curtis-Lake}, {Egami}, {Eisenstein}, {Ji}, {Maiolino}, {Sandles}, {Smit}, {{\"U}bler}, \& {Willott}}]{Simmonds23}
{Simmonds}, C., {Tacchella}, S., {Maseda}, M., {et~al.} 2023, \mnras, 523, 5468

\bibitem[{{Smith} {et~al.}(2022){Smith}, {Kannan}, {Garaldi}, {Vogelsberger}, {Pakmor}, {Springel}, \& {Hernquist}}]{Smith22}
{Smith}, A., {Kannan}, R., {Garaldi}, E., {et~al.} 2022, \mnras, 512, 3243

\bibitem[{{Sobral} \& {Matthee}(2019)}]{Sobral19}
{Sobral}, D., \& {Matthee}, J. 2019, \aap, 623, A157

\bibitem[{{Songaila} {et~al.}(2022){Songaila}, {Barger}, {Cowie}, {Hu}, \& {Taylor}}]{Songaila22}
{Songaila}, A., {Barger}, A.~J., {Cowie}, L.~L., {Hu}, E.~M., \& {Taylor}, A.~J. 2022, \apj, 935, 52

\bibitem[{{Songaila} {et~al.}(2024){Songaila}, {Cowie}, {Barger}, {Hu}, \& {Taylor}}]{Songaila24}
{Songaila}, A., {Cowie}, L.~L., {Barger}, A.~J., {Hu}, E.~M., \& {Taylor}, A.~J. 2024, arXiv e-prints, arXiv:2407.08772

\bibitem[{{Songaila} {et~al.}(2018){Songaila}, {Hu}, {Barger}, {Cowie}, {Hasinger}, {Rosenwasser}, \& {Waters}}]{Songaila18}
{Songaila}, A., {Hu}, E.~M., {Barger}, A.~J., {et~al.} 2018, \apj, 859, 91

\bibitem[{{Stark} {et~al.}(2017){Stark}, {Ellis}, {Charlot}, {Chevallard}, {Tang}, {Belli}, {Zitrin}, {Mainali}, {Gutkin}, {Vidal-Garc{\'\i}a}, {Bouwens}, \& {Oesch}}]{Stark17}
{Stark}, D.~P., {Ellis}, R.~S., {Charlot}, S., {et~al.} 2017, \mnras, 464, 469

\bibitem[{{Steidel} {et~al.}(2018){Steidel}, {Bogosavljevi{\'c}}, {Shapley}, {Reddy}, {Rudie}, {Pettini}, {Trainor}, \& {Strom}}]{Steidel18}
{Steidel}, C.~C., {Bogosavljevi{\'c}}, M., {Shapley}, A.~E., {et~al.} 2018, \apj, 869, 123

\bibitem[{{Tacchella} {et~al.}(2018){Tacchella}, {Bose}, {Conroy}, {Eisenstein}, \& {Johnson}}]{Tacchella18}
{Tacchella}, S., {Bose}, S., {Conroy}, C., {Eisenstein}, D.~J., \& {Johnson}, B.~D. 2018, \apj, 868, 92

\bibitem[{{Tang} {et~al.}(2024{\natexlab{a}}){Tang}, {Stark}, {Topping}, {Mason}, \& {Ellis}}]{Tang24}
{Tang}, M., {Stark}, D.~P., {Topping}, M.~W., {Mason}, C., \& {Ellis}, R.~S. 2024{\natexlab{a}}, arXiv e-prints, arXiv:2408.01507

\bibitem[{{Tang} {et~al.}(2023){Tang}, {Stark}, {Chen}, {Mason}, {Topping}, {Endsley}, {Senchyna}, {Plat}, {Lu}, {Whitler}, {Robertson}, \& {Charlot}}]{Tang23}
{Tang}, M., {Stark}, D.~P., {Chen}, Z., {et~al.} 2023, \mnras, 526, 1657

\bibitem[{{Tang} {et~al.}(2024{\natexlab{b}}){Tang}, {Stark}, {Ellis}, {Sun}, {Topping}, {Robertson}, {Tacchella}, {Arribas}, {Baker}, {Bhatawdekar}, {Boyett}, {Bunker}, {Charlot}, {Chen}, {Chevallard}, {Jones}, {Kumari}, {Lyu}, {Maiolino}, {Maseda}, {Saxena}, {Whitler}, {Williams}, {Willott}, \& {Witstok}}]{Tang24b}
{Tang}, M., {Stark}, D.~P., {Ellis}, R.~S., {et~al.} 2024{\natexlab{b}}, \mnras, 531, 2701

\bibitem[{{Taylor} {et~al.}(2020){Taylor}, {Barger}, {Cowie}, {Hu}, \& {Songaila}}]{Taylor20}
{Taylor}, A.~J., {Barger}, A.~J., {Cowie}, L.~L., {Hu}, E.~M., \& {Songaila}, A. 2020, \apj, 895, 132

\bibitem[{{Taylor} {et~al.}(2021){Taylor}, {Cowie}, {Barger}, {Hu}, \& {Songaila}}]{Taylor21}
{Taylor}, A.~J., {Cowie}, L.~L., {Barger}, A.~J., {Hu}, E.~M., \& {Songaila}, A. 2021, \apj, 914, 79

\bibitem[{{Tilvi} {et~al.}(2014){Tilvi}, {Papovich}, {Finkelstein}, {Long}, {Song}, {Dickinson}, {Ferguson}, {Koekemoer}, {Giavalisco}, \& {Mobasher}}]{Tilvi14}
{Tilvi}, V., {Papovich}, C., {Finkelstein}, S.~L., {et~al.} 2014, \apj, 794, 5

\bibitem[{{Tilvi} {et~al.}(2020){Tilvi}, {Malhotra}, {Rhoads}, {Coughlin}, {Zheng}, {Finkelstein}, {Veilleux}, {Mobasher}, {Wang}, {Probst}, {Swaters}, {Hibon}, {Joshi}, {Zabl}, {Jiang}, {Pharo}, \& {Yang}}]{Tilvi20}
{Tilvi}, V., {Malhotra}, S., {Rhoads}, J.~E., {et~al.} 2020, \apjl, 891, L10

\bibitem[{{Torralba-Torregrosa} {et~al.}(2024){Torralba-Torregrosa}, {Matthee}, {Naidu}, {Mackenzie}, {Pezzulli}, {Hutter}, {Arnalte-Mur}, {Gurung-L{\'o}pez}, {Tacchella}, {Oesch}, {Kashino}, {Conroy}, \& {Sobral}}]{Torregrosa24}
{Torralba-Torregrosa}, A., {Matthee}, J., {Naidu}, R.~P., {et~al.} 2024, arXiv e-prints, arXiv:2404.10040

\bibitem[{{Umeda} {et~al.}(2023){Umeda}, {Ouchi}, {Nakajima}, {Harikane}, {Ono}, {Xu}, {Isobe}, \& {Zhang}}]{Umeda23}
{Umeda}, H., {Ouchi}, M., {Nakajima}, K., {et~al.} 2023, arXiv e-prints, arXiv:2306.00487

\bibitem[{{Urrutia} {et~al.}(2019){Urrutia}, {Wisotzki}, {Kerutt}, {Schmidt}, {Herenz}, {Klar}, {Saust}, {Werhahn}, {Diener}, {Caruana}, {Krajnovi{\'c}}, {Bacon}, {Boogaard}, {Brinchmann}, {Enke}, {Maseda}, {Nanayakkara}, {Richard}, {Steinmetz}, \& {Weilbacher}}]{Urrutia19}
{Urrutia}, T., {Wisotzki}, L., {Kerutt}, J., {et~al.} 2019, \aap, 624, A141

\bibitem[{{Valentino} {et~al.}(2022){Valentino}, {Brammer}, {Fujimoto}, {Heintz}, {Weaver}, {Strait}, {Gould}, {Mason}, {Watson}, {Laursen}, \& {Toft}}]{Valentino22}
{Valentino}, F., {Brammer}, G., {Fujimoto}, S., {et~al.} 2022, \apjl, 929, L9

\bibitem[{{Verhamme} {et~al.}(2015){Verhamme}, {Orlitov{\'a}}, {Schaerer}, \& {Hayes}}]{Verhamme15}
{Verhamme}, A., {Orlitov{\'a}}, I., {Schaerer}, D., \& {Hayes}, M. 2015, \aap, 578, A7

\bibitem[{{Verhamme} {et~al.}(2017){Verhamme}, {Orlitov{\'a}}, {Schaerer}, {Izotov}, {Worseck}, {Thuan}, \& {Guseva}}]{Verhamme17}
{Verhamme}, A., {Orlitov{\'a}}, I., {Schaerer}, D., {et~al.} 2017, \aap, 597, A13

\bibitem[{{Verhamme} {et~al.}(2018){Verhamme}, {Garel}, {Ventou}, {Contini}, {Bouch{\'e}}, {Herenz}, {Richard}, {Bacon}, {Schmidt}, {Maseda}, {Marino}, {Brinchmann}, {Cantalupo}, {Caruana}, {Cl{\'e}ment}, {Diener}, {Drake}, {Hashimoto}, {Inami}, {Kerutt}, {Kollatschny}, {Leclercq}, {Patr{\'\i}cio}, {Schaye}, {Wisotzki}, \& {Zabl}}]{Verhamme18}
{Verhamme}, A., {Garel}, T., {Ventou}, E., {et~al.} 2018, \mnras, 478, L60

\bibitem[{{Weilbacher} {et~al.}(2020){Weilbacher}, {Palsa}, {Streicher}, {Bacon}, {Urrutia}, {Wisotzki}, {Conseil}, {Husemann}, {Jarno}, {Kelz}, {P{\'e}contal-Rousset}, {Richard}, {Roth}, {Selman}, \& {Vernet}}]{Weilbacher20}
{Weilbacher}, P.~M., {Palsa}, R., {Streicher}, O., {et~al.} 2020, \aap, 641, A28

\bibitem[{{Weinberger} {et~al.}(2018){Weinberger}, {Kulkarni}, {Haehnelt}, {Choudhury}, \& {Puchwein}}]{Weinberger18}
{Weinberger}, L.~H., {Kulkarni}, G., {Haehnelt}, M.~G., {Choudhury}, T.~R., \& {Puchwein}, E. 2018, \mnras, 479, 2564

\bibitem[{{Whitler} {et~al.}(2023){Whitler}, {Endsley}, {Stark}, {Topping}, {Chen}, \& {Charlot}}]{Whitler23}
{Whitler}, L., {Endsley}, R., {Stark}, D.~P., {et~al.} 2023, \mnras, 519, 157

\bibitem[{{Whitler} {et~al.}(2024){Whitler}, {Stark}, {Endsley}, {Chen}, {Mason}, {Topping}, \& {Charlot}}]{Whitler24}
{Whitler}, L., {Stark}, D.~P., {Endsley}, R., {et~al.} 2024, \mnras, 529, 855

\bibitem[{{Whitler} {et~al.}(2020){Whitler}, {Mason}, {Ren}, {Dijkstra}, {Mesinger}, {Pentericci}, {Trenti}, \& {Treu}}]{Whitler20}
{Whitler}, L.~R., {Mason}, C.~A., {Ren}, K., {et~al.} 2020, \mnras, 495, 3602

\bibitem[{{Witstok} {et~al.}(2024){Witstok}, {Smit}, {Saxena}, {Jones}, {Helton}, {Sun}, {Maiolino}, {Kumari}, {Stark}, {Bunker}, {Arribas}, {Baker}, {Bhatawdekar}, {Boyett}, {Cameron}, {Carniani}, {Charlot}, {Chevallard}, {Curti}, {Curtis-Lake}, {Eisenstein}, {Endsley}, {Hainline}, {Ji}, {Johnson}, {Looser}, {Nelson}, {Perna}, {Rix}, {Robertson}, {Sandles}, {Scholtz}, {Simmonds}, {Tacchella}, {{\"U}bler}, {Williams}, {Willmer}, \& {Willott}}]{Witstok24}
{Witstok}, J., {Smit}, R., {Saxena}, A., {et~al.} 2024, \aap, 682, A40

\bibitem[{{Witten} {et~al.}(2024){Witten}, {Laporte}, {Martin-Alvarez}, {Sijacki}, {Yuan}, {Haehnelt}, {Baker}, {Dunlop}, {Ellis}, {Grogin}, {Illingworth}, {Katz}, {Koekemoer}, {Magee}, {Maiolino}, {McClymont}, {P{\'e}rez-Gonz{\'a}lez}, {Pusk{\'a}s}, {Roberts-Borsani}, {Santini}, \& {Simmonds}}]{Witten24}
{Witten}, C., {Laporte}, N., {Martin-Alvarez}, S., {et~al.} 2024, Nature Astronomy, 8, 384

\bibitem[{{Witten} {et~al.}(2023){Witten}, {Laporte}, \& {Katz}}]{Witten23}
{Witten}, C. E.~C., {Laporte}, N., \& {Katz}, H. 2023, \apj, 944, 61

\bibitem[{{Xu} {et~al.}(2022){Xu}, {Henry}, {Heckman}, {Chisholm}, {Worseck}, {Gronke}, {Jaskot}, {McCandliss}, {Flury}, {Giavalisco}, {Ji}, {Amor{\'\i}n}, {Berg}, {Borthakur}, {Bouche}, {Carr}, {Erb}, {Ferguson}, {Garel}, {Hayes}, {Makan}, {Marques-Chaves}, {Rutkowski}, {{\"O}stlin}, {Rafelski}, {Saldana-Lopez}, {Scarlata}, {Schaerer}, {Trebitsch}, {Tremonti}, {Verhamme}, \& {Wang}}]{Xu22}
{Xu}, X., {Henry}, A., {Heckman}, T., {et~al.} 2022, \apj, 933, 202

\bibitem[{{Yajima} {et~al.}(2018){Yajima}, {Sugimura}, \& {Hasegawa}}]{Yajima18}
{Yajima}, H., {Sugimura}, K., \& {Hasegawa}, K. 2018, \mnras, 477, 5406

\bibitem[{{Yang} {et~al.}(2017){Yang}, {Malhotra}, {Gronke}, {Rhoads}, {Leitherer}, {Wofford}, {Jiang}, {Dijkstra}, {Tilvi}, \& {Wang}}]{Yang17}
{Yang}, H., {Malhotra}, S., {Gronke}, M., {et~al.} 2017, \apj, 844, 171

\end{thebibliography}
%\printbibliography

%\newpage
\appendix
\section{Fitting \lya\ lines}\label{app1}
\vspace{2mm}
\begin{figure}[h!]
    \centering
    \includegraphics[width=5.8cm,height=3.9cm]{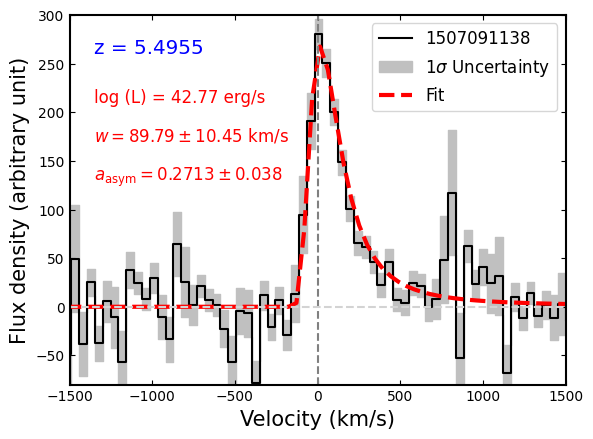}
    \includegraphics[width=5.8cm,height=3.9cm]{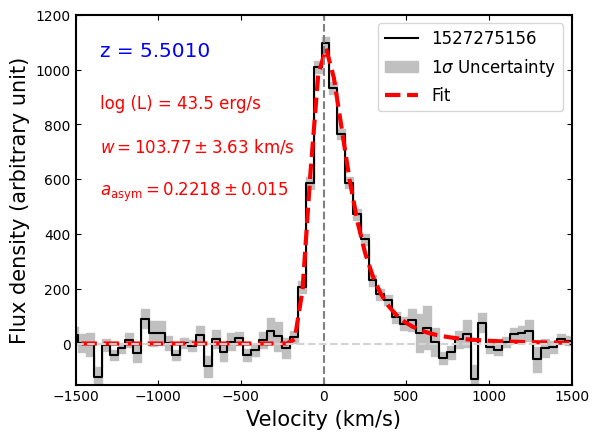}
    \includegraphics[width=5.8cm,height=3.9cm]{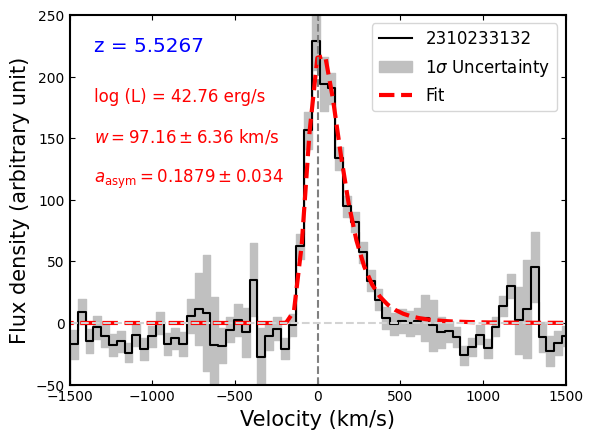}
\end{figure}

\begin{figure*}[ht!]
\centering
%    {\includegraphics[width=5.2cm,height=3.8cm]{Images/1507091138fit.png}}
%    {\includegraphics[width=5.2cm,height=3.8cm,clip]{Images/1527275156fit.png}}
%    {\includegraphics[width=5.2cm,height=3.8cm,clip]{Images/2310233132fit.png}}
%    \medskip
%    {\includegraphics[width=5.2cm,height=3.8cm]{Images/1527282124fit.png}}
    {\includegraphics[width=5.8cm,height=4cm]{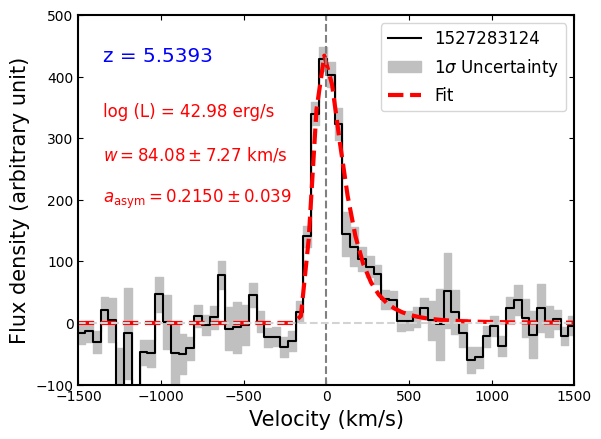}}
    {\includegraphics[width=5.8cm,height=4cm,clip]{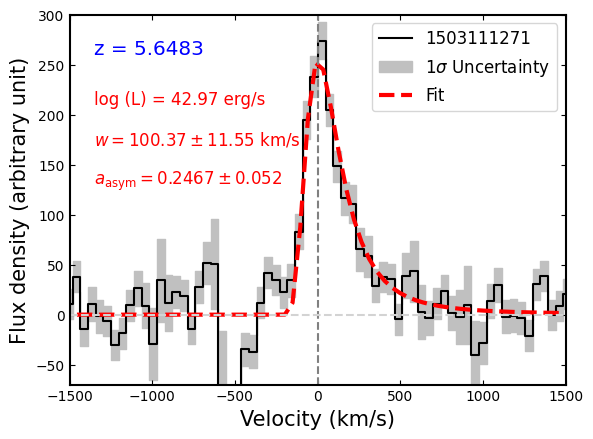}}
    {\includegraphics[width=5.8cm,height=4cm,clip]{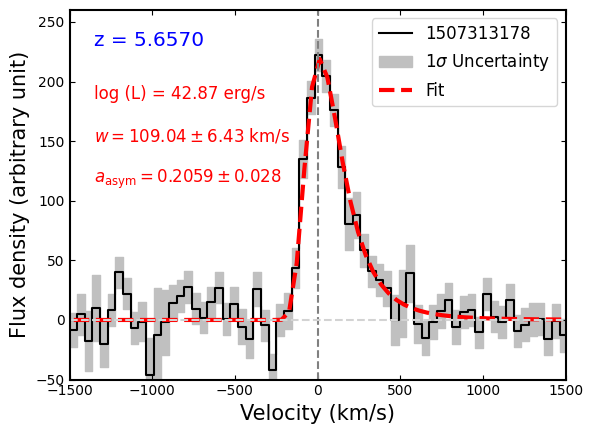}}
   
    \medskip   
     {\includegraphics[width=5.8cm,height=4cm]{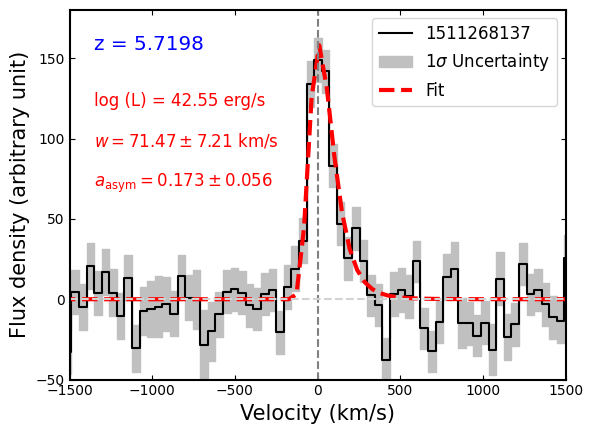}}
    {\includegraphics[width=5.8cm,height=4cm,clip]{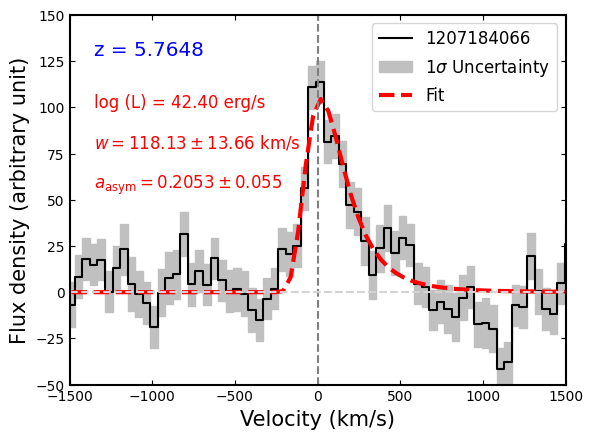}}
    {\includegraphics[width=5.8cm,height=4cm,clip]{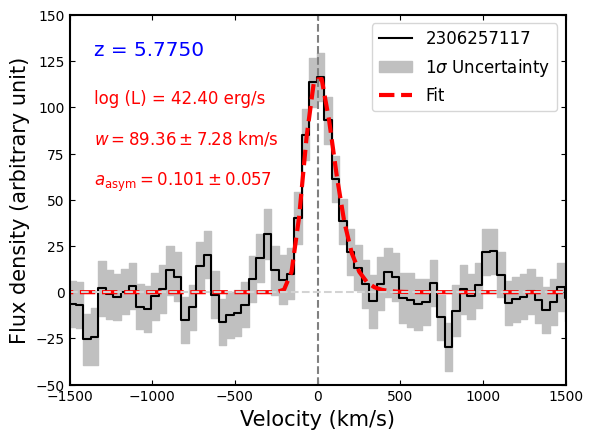}}
    \medskip
    {\includegraphics[width=5.8cm,height=4cm]{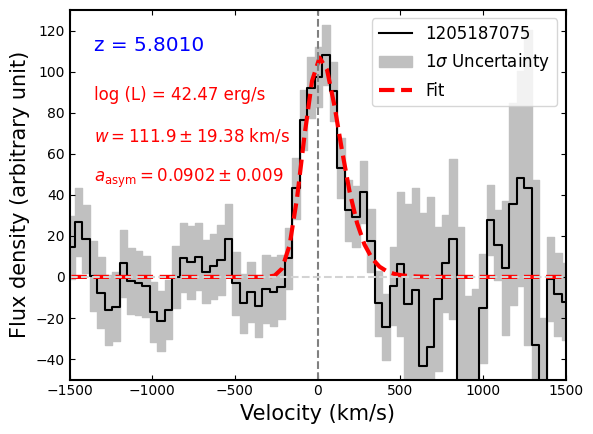}}
    {\includegraphics[width=5.8cm,height=4cm,clip]{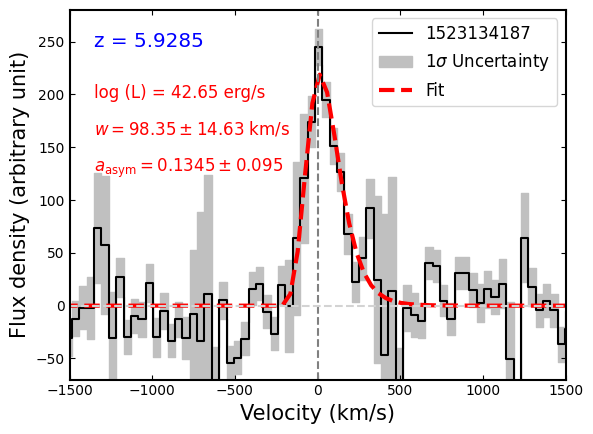}}
    {\includegraphics[width=5.8cm,height=4cm,clip]{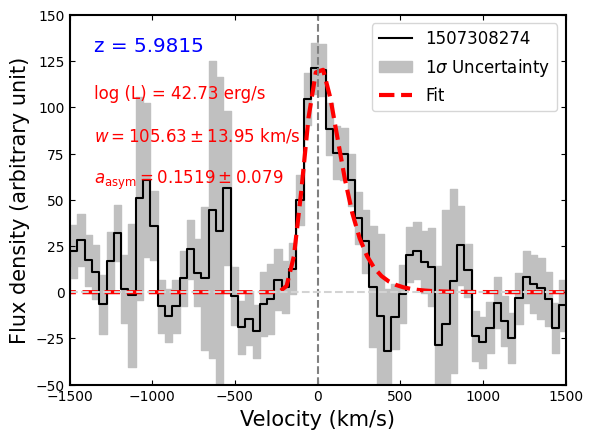}}
    \medskip    
    {\includegraphics[width=5.8cm,height=4cm]{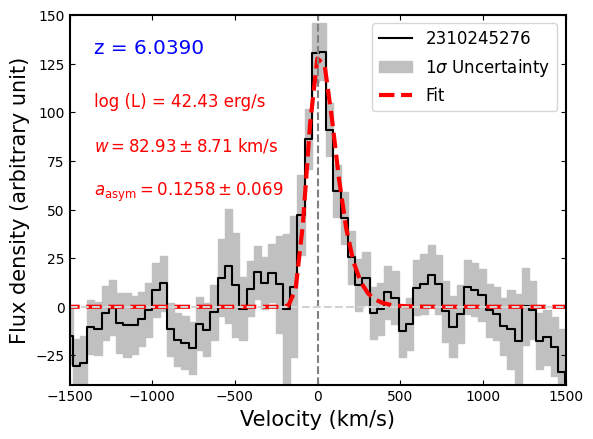}}
    {\includegraphics[width=5.8cm,height=4cm,clip]{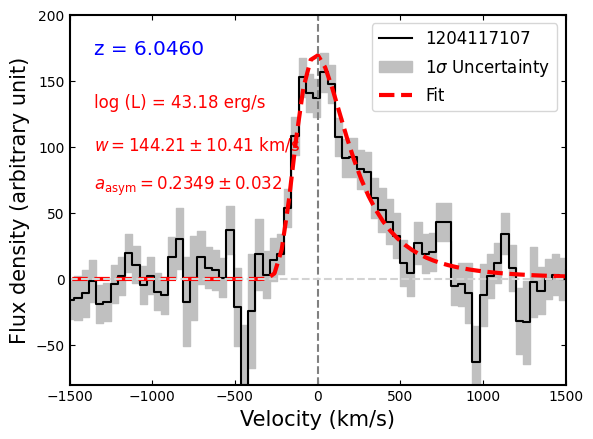}}
    {\includegraphics[width=5.8cm,height=4cm]{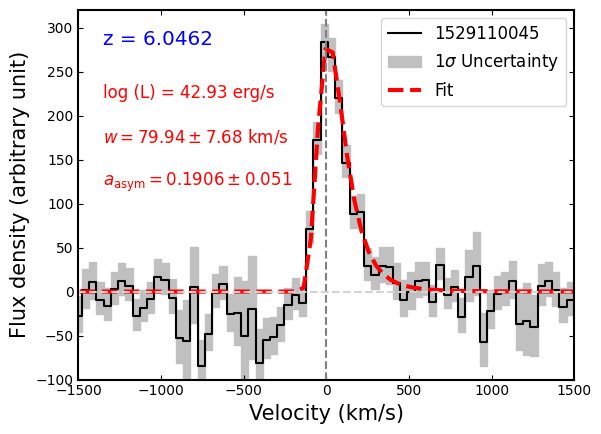}}
    \medskip
    {\includegraphics[width=5.8cm,height=4cm]{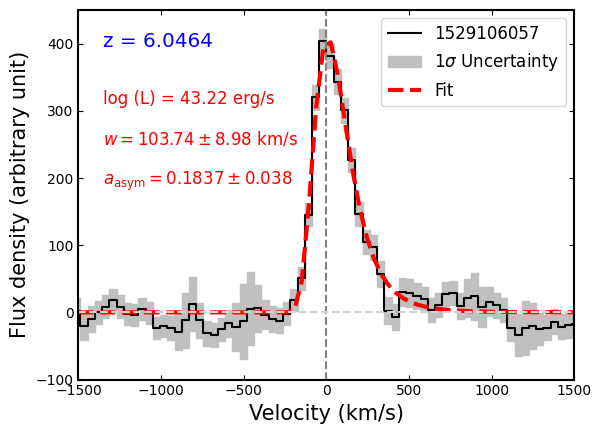}}
    {\includegraphics[width=5.8cm,height=4cm,clip]{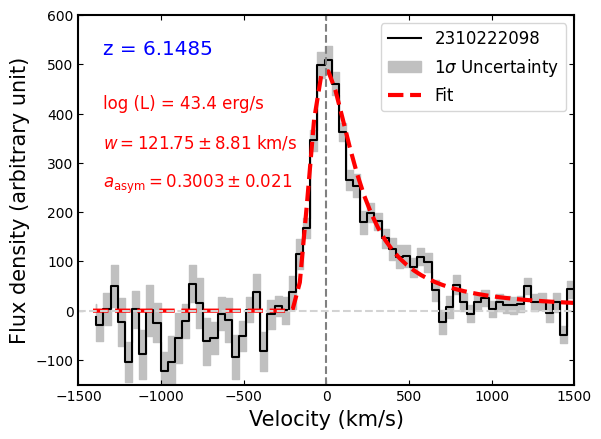}}
    {\includegraphics[width=5.8cm,height=4cm,clip]{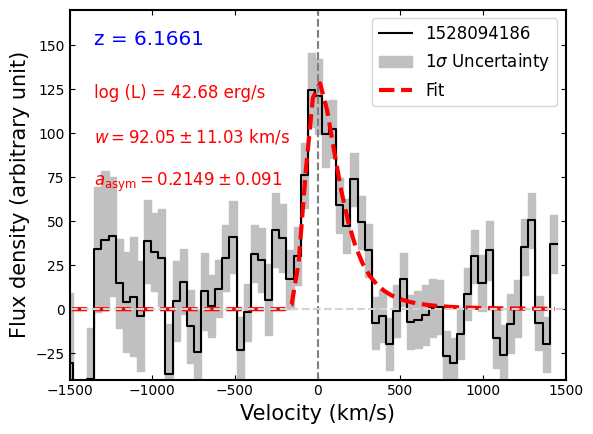}} 
      
%    \medskip    
%    {\includegraphics[width=5.2cm,height=3.8cm]{Images/1529106057fit.png}}
%    {\includegraphics[width=5.2cm,height=3.8cm,clip]{Images/2310222098fit.png}}
%    {\includegraphics[width=5.2cm,height=3.8cm,clip]{Images/1528094186fit.png}}

%\caption{Shown are velocity profiles of LAEs within $5.5 \lesssim z < 6$. 1D spectrum (solid black), error spectrum (grey shaded) together with the corresponding asymmetric Gaussian fit (dashed-red) for each LAE are presented in separate panels. The LAE MAGPI ID is provided in the legend. In each panel, luminosity, and best-fit parameters such as redshift, $w$ defining FWHM, and asymmetry parameter are given in the left corners.  }
\label{a1}
\end{figure*}

\begin{figure*}[ht!]
%\centering
    {\includegraphics[width=5.8cm,height=4cm,clip]{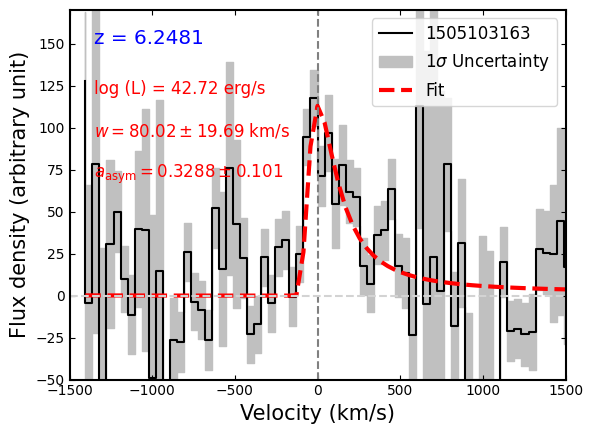}}
    {\includegraphics[width=5.8cm,height=4cm]{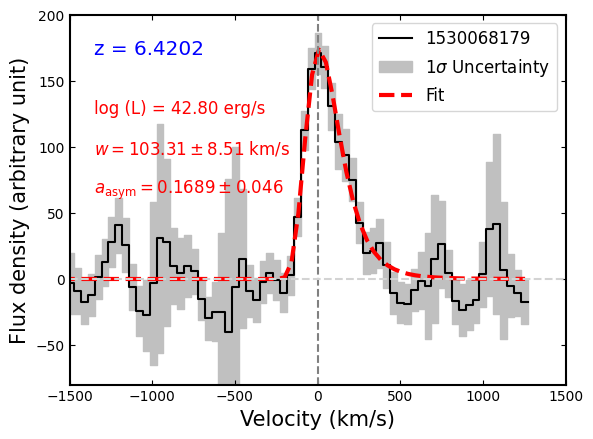}}    {\includegraphics[width=5.8cm,height=4cm,clip]{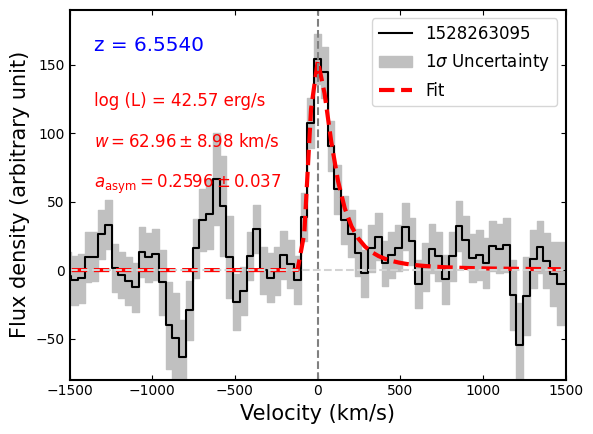}}
    \medskip
    {\includegraphics[width=5.8cm,height=4cm,clip]{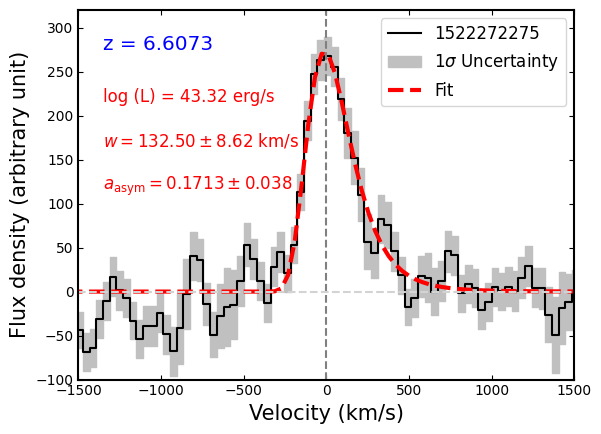}}
%    \medskip
    
\caption{Shown are the velocity profiles of $22$ LAEs at $5.5 \lesssim z \lesssim 6.6$. 1D spectrum (solid black, named by corresponding MAGPI ID) and associated $\pm 1 \sigma$ uncertainty on the flux density (grey shaded), along with the asymmetric Gaussian fit (dashed-red) to the spectrum for each LAE are presented in each panels. In each panel, luminosity, redshift and best-fit parameters such as $w$ defining FWHM, and asymmetry parameter are given in the left corners. }
\label{a2}
\end{figure*}

%\section{Example Appendix Section}
%Lorem ipsum dolor sit amet, consectetur adipiscing elit, sed do eiusmod tempor incididunt ut labore et dolore magna aliqua. Lorem ipsum dolor sit amet, consectetur adipiscing elit, sed do eiusmod tempor incididunt ut labore et dolore magna aliqua. Lorem ipsum dolor sit amet, consectetur adipiscing elit, sed do eiusmod tempor incididunt ut labore et dolore magna aliqua. 

\end{document}